\documentclass{emulateapj}
\usepackage{amssymb,latexsym,amsmath,graphicx,rotating,hyperref,longtable}
\usepackage{natbib}
\shorttitle{SMBH Growth in Starburst Galaxies}
\shortauthors{Rafferty et al.}
\slugcomment{Accepted to The Astrophysical Journal}

\newcommand{\um}{$\mu$m}

\begin{document}

\setlength{\pdfpageheight}{\paperheight}
\setlength{\pdfpagewidth}{\paperwidth}

\title{Supermassive Black Hole Growth in Starburst Galaxies over Cosmic Time:\\ Constraints from the Deepest \emph{Chandra} Fields}
\author{D.~A.~Rafferty\altaffilmark{1,2}, W.~N.~Brandt\altaffilmark{1}, D.~M.~Alexander\altaffilmark{3}, Y.~Q.~Xue\altaffilmark{1}, F.~E.~Bauer\altaffilmark{4,5}, \\B.~D.~Lehmer\altaffilmark{3,6,7}, B.~Luo\altaffilmark{1}, and C.~Papovich\altaffilmark{8}}

\altaffiltext{1}{Department of Astronomy and Astrophysics, Pennsylvania State University, University Park, PA 16802, USA}
\altaffiltext{2}{Sterrewacht Leiden, Leiden University, P.O. Box 9513, 2300 RA, Leiden, The Netherlands}
\altaffiltext{3}{Department of Physics, Durham University, Durham, DH1 3LE, UK}
\altaffiltext{4}{Space Science Institute, 4750 Walnut Street, Suite
205, Boulder, Colorado 80301}
\altaffiltext{5}{Pontificia Universidad Cat\'{o}lica de Chile,
Departamento de Astronom\'{\i}a y Astrof\'{\i}sica, Casilla 306,
Santiago 22, Chile}
\altaffiltext{6}{The Johns Hopkins University, Homewood Campus, Baltimore, MD 21218, USA}
\altaffiltext{7}{NASA Goddard Space Flight Centre, Code 662, Greenbelt, MD 20771, USA}
\altaffiltext{8}{Department of Physics, Texas A\&M University, 4242 TAMU, College Station, TX 77843, USA}

\begin{abstract}
We present an analysis of deep multiwavelength data for $z\approx0.3$--3 starburst galaxies selected by their 70~\um\ emission in the Extended-\emph{Chandra} Deep Field-South and Extended Groth Strip. We identify active galactic nuclei (AGNs) in these infrared sources through their \mbox{X-ray} emission and quantify the fraction that host an AGN. We find that the fraction depends strongly on both the mid-infrared color and rest-frame mid-infrared luminosity of the source, rising to $\sim 50$--70\% at the warmest colors ($F_{\rm 24 \mu m}/F_{\rm 70 \mu m} \lesssim 0.2$) and highest mid-infrared luminosities (corresponding to ultraluminous infrared galaxies), similar to the trends found locally. Additionally, we find that the AGN fraction depends strongly on the star formation rate of the host galaxy (inferred from the observed-frame 70~\um\ luminosity after subtracting the estimated AGN contribution), particularly for more luminous AGNs ($L_{0.5-8.0\rm{~keV}} \gtrsim 10^{43}$~erg~s$^{-1}$). At the highest star formation rates ($\sim 1000$ M$_{\odot}$ yr$^{-1}$), the fraction of galaxies with an \mbox{X-ray} detected AGN rises to $\approx 30$\%, roughly consistent with that found in high-redshift submillimeter galaxies. Assuming that the AGN fraction is driven by the star formation rate (rather than stellar mass or redshift, for which our sample is largely degenerate), this result implies that the duty cycle of luminous AGN activity increases with the star formation rate of the host galaxy: specifically, we find that luminous \mbox{X-ray} detected AGNs are at least $\sim 5$--10 times more common in systems with high star formation rates ($\gtrsim 300$ M$_{\odot}$ yr$^{-1}$) than in systems with lower star formation rates ($\lesssim 30$ M$_{\odot}$ yr$^{-1}$). Lastly, we investigate the ratio between the supermassive black hole accretion rate (inferred from the AGN \mbox{X-ray} luminosity) and the bulge growth rate of the host galaxy (approximated as the star formation rate) and find that, for sources with detected AGNs and star formation (and neglecting systems with low star formation rates to which our data are insensitive), this ratio in distant starbursts agrees well with that expected from the local scaling relation assuming the black holes and bulges grew at the same epoch. These results imply that black holes and bulges grow together during periods of vigorous star formation and AGN activity.
\end{abstract}

\section{Introduction}
The observed scaling between the mass of a galaxy's bulge and the mass of its central supermassive black hole (SMBH) points to a fundamental connection between the growth of galaxies and their BHs. Recent findings \citep[e.g.,][]{mago98,ferr00,gebh00,alex05b,hopk06,wild07} suggest that SMBHs and bulges generally grow together; however, many of the details are still unclear, particularly at intermediate and high redshift \citep[e.g.,][]{shie06,alex08_bh}. The signatures of SMBH and bulge growth are both readily observable over a broad redshift range, as SMBH growth produces an AGN and bulge growth is accompanied by active star formation.  A simple observable that links these two indicators is the AGN fraction as a function of star formation rate. A determination of the form of this relation over a broad range of redshift would provide new constraints on large-scale models of galaxy formation and evolution \citep[e.g.,][]{dima08,hopk08,youn09}. 

The AGN fraction, for a given sample, is the number of systems with AGN activity divided by the total number of systems in which such activity could have been detected (e.g., to some AGN luminosity limit), given the sensitivity limits of the observations. The AGN fraction provides clues to the duty cycle of SMBH accretion: a higher fraction implies that the SMBHs spend less time in inactive states relative to that spent in active accreting states. Therefore, any dependence of the AGN fraction on star formation rate would imply that this duty cycle is related to the intensity of star formation. In particular, studies that identify AGNs using optical spectra have shown that the luminous AGN fraction in all galaxies at $z\approx 0$ is on the order of 5--15\% percent \citep[e.g.,][]{kauf03,fran04}, whereas in massively star forming galaxies, such as the sub-mm galaxies (SMGs) studied by \citet{alex05b} and \citet{lair10} at $z\approx 2$, the fraction is estimated to be considerably higher: Alexander et al.\ find a fraction of $38^{+12}_{-10}$\% using \mbox{X-ray} and radio data, and Laird et al.\ derive a somewhat lower fraction of (20--29)$\pm7$\% using \mbox{X-ray}-selected AGNs. Although these numbers agree within errors, their factor of $\sim 2$ difference points to the need for additional studies of the AGN fraction in this high SFR regime, using larger samples and different approaches. Additionally, the detailed form of the AGN fraction between the two extremes of star formation rate is currently poorly constrained.

Due to dust that absorbs the UV emission from young stars and re-emits it at long wavelengths, a galaxy's mid-to-far-infrared emission is commonly used as a tracer of its star formation activity.\footnote{In this paper, we define mid-infrared emission to be emission at rest-frame wavelengths $\lesssim 40$~\um, and far-infrared emission to be at rest-frame wavelengths $\gtrsim 40$~\um.} Recently, very deep \textit{Spitzer} MIPS data have become available for the deepest \textit{Chandra} \mbox{\mbox{X-ray}} fields, which together provide sensitive \mbox{X-ray} and mid-to-far-infrared observations that are ideal for identifying AGN activity to luminosities of $L_{\rm bol} \sim 10^{42}$~erg~s$^{-1}$ and dust-obscured star formation to star-formation rates of $\sim 10$~M$_{\odot}$ yr$^{-1}$ at $z\sim 0.5$. These data allow one to trace luminous star formation and AGN activity in the distant universe and to investigate how the AGN fraction depends on the star formation rate. 

However, it is well known that AGNs are associated with dusty ``tori'' that are often luminous in the mid-infrared. Therefore, AGN may also contribute significantly to the total infrared emission when present. A number of studies have investigated the contribution from AGN-powered emission to the infrared flux in luminous infrared sources. Such sources are generally divided into two subclasses by their integrated 8--1000~\um\ luminosity (denoted $L_{\rm IR}$): luminous infrared galaxies (LIRGs, $ 10^{11} < L_{\rm IR} < 10^{12}$~$L_{\odot}$) and ultraluminous infrared galaxies (ULIRGs, $L_{\rm IR} > 10^{12}$~$L_{\odot}$). Among the general population of luminous infrared sources, star formation appears to be the dominant power source of the mid-to-far-infrared emission in most objects. For example, using diagnostics based on mid-infrared emission lines and the strength of the 7.7~\um\ PAH feature of $z\lesssim 0.15$ ULIRGS, \citet{genz98} found that star formation likely powers most of the 8--1000~\um\ luminosity of 12 of the 15 ULIRGs they studied. Using similar mid-infrared diagnostics, \citet{houc07} found that for the majority of their sample, selected by 24~\um\ flux and mostly at $z\lesssim 0.25$, the mid-infrared luminosity is dominated by star formation. In a sample of 43 $0.1 < z < 1.2$ objects selected by 70~\um\ flux, \citet{syme08} fit a variety of starburst and AGN-powered emission models to IRAC and MIPS photometry and IRS spectra and found that all but one object in their sample are starburst dominated. 

In a study of high-luminosity systems at $z<0.5$, \citet{tran01} found that at luminosities below $L_{\rm IR} \sim 10^{12.5}$~$L_{\odot}$, starbursts (identified in ISO spectra by their strong mid-infrared PAH emission) are the dominant power source in local ULIRGs, but at higher luminosities, AGN are often the dominant emission source. However, in a study of \emph{Spitzer} IRS spectra of a sample of 107 ULIRGs, \citet{desa07} found that even the most luminous high-redshift ULIRGs often have strong PAH emission, indicative of large starbursts that may be absent locally. Lastly, the recent study of \citet{veil09}, which also used IRS spectra of ULIRGs to estimate the relative contributions of AGNs and starbursts, found that the average AGN contribution to the bolometric (not far-infrared) luminosity of local ($z\sim 0.3$) ULIRGs is $\sim 35$--40\%. However, among far-infrared sources with the most luminous AGNs, namely quasars, AGN-powered emission can dominate. For example, \citet{shi07} used the mid-infrared PAH emission to infer SFRs in three samples of AGNs: PG quasars, 2MASS quasars, and 3CR radio-loud AGNs. They found that the average contribution of star formation to the 70~\um\ emission ranges from 25\%--50\%, depending on the AGN sample. Therefore, one must be careful to account for the AGN contribution to the infrared when using it to derive star formation rates.

In this paper, we investigate the growth of SMBHs and their host galaxies in a complete sample of starburst galaxies, constructed from fields with extremely deep multiwavelength coverage, and determine the \mbox{X-ray}-detected AGN fraction across a broad range of star formation rate and redshift. Briefly, our sample is constructed using the following approach:
\begin{enumerate}
\item Since mid-to-far-infrared observations sample the bulk of reprocessed emission from young stars, we use deep mid-infrared (70~\um) data to construct a representative sample of star-forming galaxies.
\item We use deep \mbox{X-ray} observations to identify AGNs above a given \mbox{X-ray} luminosity.
\item In such sources, we use a variety of empirical AGN SEDs, scaled by the AGN bolometric luminosity estimated from the \mbox{X-ray} emission, to estimate the AGN contribution to the infrared luminosity.
\item Lastly, we use the net infrared luminosity, corrected for AGN-powered emission, to estimate star formation rates.
\end{enumerate}
Using this sample, we calculate the \mbox{X-ray}-detected AGN fraction above a given limiting \mbox{X-ray} luminosity as a function of the SFR, mid-infrared color (a proxy for dust temperature), and mid-infrared luminosity, and we examine the relative growth rates of the galaxies and their SMBHs in distant starbursts. The following sections describe in detail each of these steps. We adopt $H_0 = 70$~km~s$^{-1}$~Mpc$^{-1}$, $\Omega_{\Lambda} = 0.7$, and $\Omega_{\rm{M}} = 0.3$ throughout.

\section{Sample Construction and Properties}\label{S:sample}
\subsection{Mid-Infrared Data}
Samples were drawn from two fields with deep \mbox{X-ray} through infrared coverage: the Extended-\textit{Chandra} Deep Field-South (\mbox{E-CDF-S}), which includes the $\sim 2$~Ms \textit{Chandra} Deep Field-South (\mbox{CDF-S}), and the Extended Groth Strip (EGS). The primary sample of star-forming galaxies and AGNs was constructed using all sources with \emph{Spitzer} MIPS detections at 70~\um\ in the Far-Infrared Deep Extragalactic Legacy (FIDEL) survey. The FIDEL data comprise very deep coverage of $\gtrsim 90$\% of the \mbox{E-CDF-S} and EGS fields at 24~\um\ and 70~\um. Source catalogs were created from the DR2 mosaic images\footnote{See \url{http://irsa.ipac.caltech.edu/data/SPITZER/FIDEL}.} using the DAOPHOT tool.\footnote{See \url{http://www.star.bris.ac.uk/~mbt/daophot}.} Aperture fluxes measured by DAOPHOT were corrected using the point source function derived by \citet{fray06} to derive the total fluxes. No color corrections were performed, as they are expected to be $\lesssim 10$\% for the bulk of our sources. We use the FIDEL 70~\um\ catalog as the basis of our sample because emission at this wavelength should suffer less from spectral complexity and have a smaller AGN contribution for sources at redshifts up to $\sim 3$ than emission at 24~\um. For example, at $z\gtrsim 1$, 24~\um\ observations would sample rest-frame emission at $\lambda \lesssim 12$~\um\ where complex spectral features from PAH emission and silicate absorption are present in the spectra of many LIRGS and AGNs \citep[e.g.,][]{weed05,armu07,desa07}. 

Before attempting to identify the counterparts at other wavelengths, we made a cut at a 70~\um\ signal-to-noise ratio (S/N) of 3, above which the median positional uncertainty in the 70~\um\ sources is $\lesssim 3$\arcsec\ (although it reaches $\sim 8$\arcsec\ near our adopted S/N cutoff) and completeness is high ($\approx 85$\%). The S/N and completeness of the 70~\um\ detections were estimated using simulations in which fake sources were inserted in the images and their fluxes recovered. To exclude regions of very low exposure near the survey edges, where spurious sources are more common even at ${\rm S/N} > 3$, we also made a cut at an exposure time of 1000~s (exposure times were taken from the exposure maps provided with the FIDEL catalogs). With this cut on exposure time, the total areal coverage of regions with both deep \mbox{X-ray} and mid-to-far-infrared data is $\approx 1100$~arcmin$^2$  in the \mbox{E-CDF-S} and $\approx 1400$~arcmin$^2$ in the EGS. The S/N and exposure-time cuts result in total source numbers of 567 and 725 in the \mbox{E-CDF-S} and EGS, respectively. The 70~\um\ flux limit for our sample varies with the exposure time (by a factor of up to $\sim 5$) and reaches a minimum flux density of $\approx 1.8$~mJy in the GOODS-S region of the \mbox{E-CDF-S}.

\subsection{Source Cross Matching}\label{S:crossmatch}
Due to the large point spread function of the \emph{Spitzer} MIPS instrument and to source blending, 70~\um\ source positions can be uncertain by large amounts (the simulations described above give errors of up to $\sim 8$\arcsec\ for a source with ${\rm S/N} = 3$). To minimize the number of spurious counterparts, we performed a cross match between the 70~\um\ and 24~\um\ catalogs using a probabilistic matching method \citep[described in detail in][]{luo10} that takes into account both the estimated source positional errors and their fluxes \citep[e.g.,][]{suth92,cili03}. This method tends to recover a larger fraction of true counterparts than the standard method that uses a single fixed matching radius and ignores flux information when selecting among possible counterparts. Using this method, we find that the expected fraction of spurious 24~\um-to-70~\um\ cross matches is $\lesssim 10$\%, a value $\approx 40$\% lower than that obtained with the standard method using a fixed matching radius that recovers approximately the same number of total counterparts.\footnote{The number of expected spurious matches was estimated by shifting one catalog relative to the other by 15--60\arcsec\ in 100 different directions, each time cross correlating the catalog sources. The average of the resulting number of matches was taken to be the expected number of spurious sources for a given search radius. We note however that this method likely overestimates the fraction of false matches, as we find that the typical distance to a counterpart is much less than the search radius when the cross match is performed at the real positions.} With this method, 527 \mbox{E-CDF-S} and 678 EGS 70~\um\ sources with ${\rm S/N} > 3$ matched to a 24~\um\ source. Since the 24~\um\ data are much deeper than the 70~\um\ data for these fields (by a factor of $\gtrsim 30$), we can reasonably expect that all 70~\um\ sources should have 24~\um\ counterparts. Indeed, a visual inspection of the $\approx 6$\% of 70~\um\ sources that lack 24~\um\ matches suggests that most suffer from significant blending that has distorted their shapes (and hence centroids). By requiring a 24~\um\ match for each 70~\um\ source, we eliminate these problematic sources that likely have positions and fluxes in error by large amounts. 

Additionally, $\approx 25$\% of the 70~\um\ sources with identified 24~\um\ counterparts have more than one 24~\um\ source within $\approx 4$\arcsec. Such multiple matches could imply significant blending is present in the 70~\um\ images, possibly leading to spurious cross matches and to misestimates of the 70~\um\ flux. To minimize spurious cross matches, we can use the observed distribution of 24/70~\um\ color for sources with unambiguous counterparts to select reliably the correct counterpart from multiple matches. To this end, we compared the 24/70~\um\ colors of each possible counterpart to the observed distribution of colors for 70~\um\ sources with single matches. If two counterparts had colors that differed significantly (by $> 2\sigma$) from the mean and the reliabilities determined from the probabilistic matching process were similar for the two, the counterpart with the color closer to the mean was chosen. In this way, we chose a different counterpart from that preferred by the probabilistic matching in $\approx 5$--10\% of cases (depending on the field) with multiple matches.  Additionally, in some cases with multiple nearby 24~\um\ sources, blending in the 70~\um\ images can be significant. However, we find that on average only $\approx 5$\% of the 24~\um\ sources have 70~\um\ counterparts. Given that we find multiple 24~\um\ sources for $\approx 25$\% of the 70~\um\ sources, each with an average of 2.2 24~\um\ sources within $\approx 4$\arcsec, we expect blending to have a significant effect on the measured 70~\um\ flux in only $1.2 \times 0.25 \times 0.05 \approx 1.5$\% of the 70~\um\ sources. At this level, the presence of blended sources should not affect our results significantly, and we therefore do not attempt to correct for them further.   

Using this 70~\um\ sample as the basis, we cross matched the sources with the following \mbox{X-ray}, optical, and infrared catalogs. For the \mbox{E-CDF-S}, we used the \emph{Chandra} \mbox{X-ray} catalogs of \citet[CDF-S]{luo08} and \citet[E-CDF-S]{lehm05}; the COMBO-17 \citep[Classifying Objects by Medium-Band Observations in 17 filters;][]{wolf04}, MUSYC \citep[Multiwavelength Survey by Yale-Chile;][]{gawi06}, and MUSIC \citep{graz06} optical and near-infrared catalogs; and the SIMPLE (\textit{Spitzer} IRAC/MUSYC Public Legacy survey in the Extended-\mbox{CDF-S}) mid-infrared catalog \citep{dame09}. For the EGS, we used the \emph{Chandra} \mbox{X-ray} catalogs of \citet{lair09}, the optical CFHTLS\footnote{See \url{http://www.cfht.hawaii.edu/Science/CFHLS/}.} and CFH12K \citep{coil04} catalogs, the near-infrared Palomar catalog \citep{bund06}, and the IRAC catalog of \citet{barm08}.

Since all of our 70~\um\ sources are required to have counterparts in the 24~\um\ data, we can use the more precise 24~\um\ positions (with typical uncertainties of $\sim 1$\arcsec) when matching to the optical catalogs, thereby reducing the number of likely spurious optical-to-70~\um\ matches. Additionally, since many of our sources should have radio counterparts (since radio emission is often associated with starbursts and AGNs), we can further refine the positions using the highly accurate astrometry of the radio-source catalogs available from VLA surveys of the fields (with typical source positional uncertainties of $\lesssim 0$\farcs2). To this end, we performed cross matching (again using the probabilistic matching method) between the 24~\um\ sources and the 20~cm catalogs of \citet{mill08} in the \mbox{E-CDF-S} and version 1.0 of the AEGIS20 catalog\footnote{See \url{http://www.roe.ac.uk/~rji/aegis20/}.} in the EGS. We found radio matches to 219 ($\approx 41$\%) \mbox{E-CDF-S} sources and 166 ($\approx 25$\%) EGS sources (note that the 20~cm EGS survey does not cover the entire field). Due to the low spatial density of radio sources in these surveys, we expect $\lesssim 1$\% of these matches to be spurious.

We next searched for optical counterparts to the 70~\um\ sources by 24~\um\ or radio source position. For 70~\um\ sources with an identified radio counterpart, we used a matching radius of 0\farcs5 (due to the much smaller positional uncertainties, probabilistic matching was not used). For sources without a radio counterpart, we used a radius of 1\arcsec. When multiple optical matches occur, we selected the source with the smallest separation. With these matching radii, we expect $\lesssim 7$\% of the optical-to-70~\um\ matches to be spurious. We then cross matched the \mbox{X-ray} source positions to the 24~\um\ or radio source position, using a variable matching radius that depends on the positional uncertainty of the \mbox{X-ray} source. We followed the method used in \citet{luo08} and define the matching radius as $r=1.5\Delta_{\rm X},$ where $\Delta_{\rm X}$ is the \mbox{X-ray} positional uncertainty given in the catalogs. Lastly, we used a matching radius of 0\farcs75 to identify matches to mid-infrared \emph{Spitzer} IRAC sources and a radius of 1\arcsec\ to match to UV GALEX sources. Table~\ref{T:cross_matches} summarizes the results of the cross matching. In total, the final sample comprises 1022 unique 70~\um\ sources with identified optical counterparts, of which 158 have an identified \mbox{X-ray} counterpart. 

\begin{deluxetable}{ccccccc}
\tablewidth{0pt}
\tablecaption{Number of sources and counterparts. \label{T:cross_matches}}
\tablehead{ \colhead{Field} & \colhead{70~\um} & \colhead{24~\um} & \colhead{Optical} & \colhead{\mbox{X-ray}} & \colhead{Near-IR} & \colhead{IRAC}}
\startdata
\mbox{E-CDF-S} &  564  &  527  &  449  &  81  & 386 &  442 \\
EGS     &  725 &  678  &  573  &  77 & 483 &  475 
\enddata
\end{deluxetable}

\subsection{Redshift Estimates}\label{S:redshifts}
To obtain redshift estimates for our 70~\um\ sources, we followed the process described above to cross match (using a 0\farcs5 radius) the optical source positions to all publicly available spectroscopic redshift catalogs of the \mbox{E-CDF-S} and EGS \citep[e.g.,][]{cris00,bunk03,dick04,le-f04,stan04,stro04,van-04,davi07}. When more than one spectroscopic redshift was available for a given source, the redshift of the higher quality (judged by the quality flags provided with the catalog) was used. If two spectroscopic redshifts were deemed of equal quality or the quality was unknown, the average was taken. In such cases, the difference between the two redshifts was typically $\lesssim 10$\%. In total for all three fields, 408 ($\approx 40$\%) of 1022 sources have high-quality spectroscopic redshifts determined from two or more spectral features.

Although the majority ($\approx 60$\%) of our sources lack high-quality spectroscopic redshifts, almost all have high-quality photometric data in multiple bands from near-UV to mid-infrared wavelengths. Although \citet{wolf04} produced a high-quality photometric redshift catalog for the entire \mbox{E-CDF-S}, new near- and mid-infrared data (from the MUSYC JHK and SIMPLE IRAC surveys; see Table~\ref{T:cross_matches} for the fraction of 70~\um\ sources in each field with near- and mid-infrared detections) have recently become available that should be particularly helpful in deriving accurate photometric redshifts for sources at higher redshifts ($z>1.4$), which are generally not available in the Wolf et al.\ catalog.\footnote{Recently, \citet{card10} used these and other data to derive high quality photometric redshifts for the \mbox{E-CDF-S}. We have verified that our results do not change when their redshifts are used instead.} Additionally, no photometric redshift catalog is currently publicly available for the EGS. Therefore, we can use all existing photometric data to obtain additional and improved photometric redshifts for our sample. To this end, we used the publicly available Zurich Extragalactic Bayesian Redshift Analyzer \citep[ZEBRA;][]{feld06} to derive redshifts for the sources that lack a spectroscopic redshift, except for \mbox{CDF-S} \mbox{X-ray} sources, for which we use the photometric redshifts of \citet{luo10}, derived using a very similar method but with a more sophisticated treatment of the photometry (e.g., including upper limits for non-detections). 
Details of the parameters used as input to ZEBRA to derive the photometric redshifts and estimates of the quality of the resulting redshifts are given in the Appendix. 

\begin{figure}
\plotone{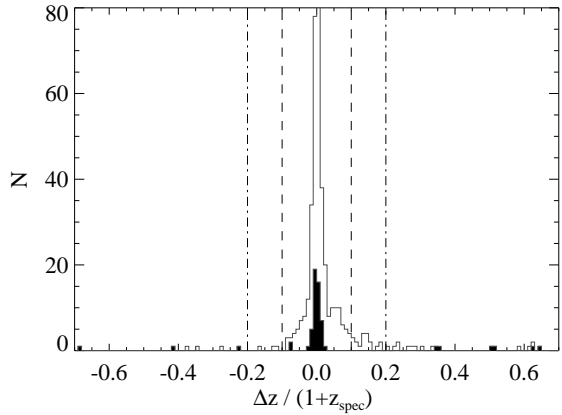}
\caption{Histograms of $\Delta z/(1+z_{\rm spec}),$ where $\Delta z = z_{\rm spec} - z_{\rm phot}$ for all 70~\um\ sources with spectroscopic redshifts (\textit{open region}) and for AGNs only (\textit{filled region}). Vertical lines denote $\Delta z/(1+z_{\rm spec})=\pm 0.1$ and $\Delta z/(1+z_{\rm spec})=\pm 0.2$.  \label{F:redshifts}}
\end{figure}

In general, the most secure photometric redshifts are those for bright sources ($m_R < 24$ AB mags), which comprise the great majority ($\approx 93$\%) of our final sample of 1022 sources. In Figure~\ref{F:redshifts}, we compare the photometric redshifts derived by ZEBRA to spectroscopic ones for the subsample of 408 sources in our final sample with spectroscopic redshifts. Assuming the spectroscopic redshifts are accurate and the spectroscopic sources are representative of the entire sample, $\approx 97$\% of our sources will have redshifts with $|z_{\rm true}-z_{\rm }|/(1+z_{\rm true}) < 0.2$. For AGNs only (identified following \S\ref{S:AGN_selection}), of which 55 of 108 have spectroscopic redshifts, the fraction of such sources is $\approx 94$\%. However, the spectroscopic sample is unlikely to be fully representative, and we therefore expect that the true fraction of sources with incorrect redshifts will be higher by roughly a factor of three (see the Appendix for details), but still within acceptable limits ($\lesssim 10$\%). 

Lastly, it should be noted that an important consequence of obtaining redshift estimates for all of the 70~\um\ sources in our sample with identified optical counterparts is that our study will not suffer from strong biases related to incomplete redshift coverage. For example, an incomplete sample based on redshift surveys that target AGNs or \mbox{X-ray} sources preferentially \citep[e.g.,][]{zhen04} would result in an artificially high AGN fraction.

\subsection{Mid-Infrared Luminosities and Colors}\label{S:midIR_lumin_color}
\begin{figure}
\plotone{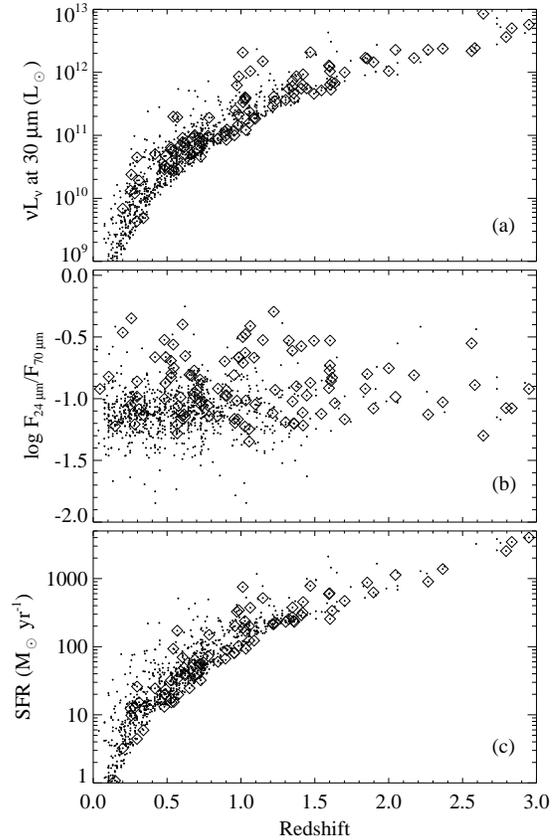}
\caption{The distributions of rest-frame 30~\um\ luminosity (\emph{a}), mid-infrared color (\emph{b}), and star formation rate (\emph{c}) of the 70~\um\ sample as a function of redshift. AGNs (selected in \S\ref{S:AGN_selection}) are indicated by diamonds. AGNs with net 70~\um\ flux densities that fall below ${\rm S/N}=3$ (after the contribution from the AGN is subtracted) have been removed from the lower panel (a total of 21 sources; see \S\ref{S:pred_AGN_fluxes} for details).\label{F:z_dist}}
\end{figure}

The rest-frame mid-infrared luminosity of a typical starburst galaxy or AGN is dominated by reprocessed emission from dust. Such emission gives a direct measure of the strength of the star formation or AGN emission that the dust reprocesses. Therefore, it is of interest to investigate how the AGN fraction relates to this luminosity. To derive the rest-frame luminosities, we first constructed observed-frame mid-infrared SEDs from the available \emph{Spitzer} data, which span observed-frame wavelengths from 3.6~\um\ to 70~\um\ (due to the large positional uncertainties inherent to the 160~\um\ data, these data were not used). The observed SED was then shifted to the rest frame of the source using its redshift. We then used linear interpolation in log space to derive the monochromatic luminosity at a rest-frame wavelength of 30~\um\ ($L_{30}$; model SED were not used, as a variety of AGN, starburst, and hybrid sources that are difficult to model are expected to be present). This wavelength was chosen to lie within the wavelength coverage of the observed SEDs of most objects, negating the need for large extrapolations. We show the rest-frame mid-infrared luminosities as a function of redshift in Figure~\ref{F:z_dist}a. It is clear from this figure that the sensitivity limits of the FIDEL survey are such that our sample is roughly complete only for sources with $L_{30}>10^{12}$~$L_{\odot}$. Below this luminosity, the completeness varies with redshift, being $\approx 100$\% at $L_{30}>10^{11}$~$L_{\odot}$ to $z=1$ and at $L_{30} > 10^{10}$~$L_{\odot}$ to $z=0.5$.

\begin{figure}
\epsscale{1.0}
\plotone{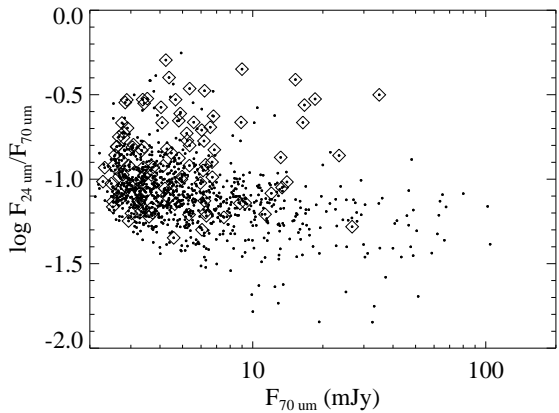}
\caption{The ratio of 24/70~\um\ flux versus the 70~\um\ flux. AGNs are indicated by diamonds. \label{F:IRcolor}}
\end{figure}

Additionally, the mid-infrared color (typically calculated using the observed-frame fluxes in the IRAS bands as $F_{\rm 25 \mu m}/F_{\rm 60 \mu m}$) of a source gives an indication of the temperature of the emitting dust: higher ratios indicate relatively more emission at shorter wavelengths, indicative of emission from warmer dust. Cooler dust temperatures are likely to be associated with dust heated by young stars, whereas warmer temperatures are more likely to be indicative of dust heated by AGN emission \citep[e.g.,][]{degr85,sand88}. Therefore, it is of interest to examine the mid-infrared colors for our sample. In Figure~\ref{F:z_dist}b, we plot the ratio of 24/70~\um\ flux (which, for our purposes, we consider to be equivalent to the ratio of 25/60~\um\ flux) against the redshift, and in Figure~\ref{F:IRcolor}, we plot it against the 70~\um\ flux. For local sources, ratios below $\log \left( F_{\rm 24 \mu m}/F_{\rm 70 \mu m} \right) \approx -0.7$ are generally indicative of emission from cool dust, whereas higher ratios are indicative of warm dust \citep[e.g.,][]{degr85,sand88}. 

It should be noted that the observed mid-infrared color for a given galaxy is a strong function of the redshift: at higher redshifts ($z\gtrsim 1.5$), the portion of the spectrum measured by observed-frame 24~\um\ emission suffers from increasing spectral complexity due to the possible presence of strong absorption and emission features below rest-frame wavelengths of $\approx 10$~\um. Additionally, there is evidence that high-redshift ($z\gtrsim 1.5$) galaxies exhibit stronger PAH emission features than local galaxies of the same luminosity \citep[e.g.,][]{murp09} that will add further redshift-dependent changes to the color. Therefore, it is difficult to infer directly a dust temperature from the mid-infrared color in sources at $z\gtrsim 1.5$ \citep[e.g.,][]{papo07,mull09}.  In \S\ref{S:fagn_sfr}, we investigate how the inclusion or exclusion of AGN hosts with warm mid-infrared colors affects our results. 

\subsection{X-ray Properties}\label{S:xray_luminosities}
The purpose of this study is to quantify the AGN fraction in mid-infrared sources; therefore, a reliable means of identifying the bulk of the AGN population is critical. Since AGNs are one of only two types of luminous \mbox{X-ray} point sources in the distant universe (the other being starburst galaxies), and \mbox{X-ray}s are not readily absorbed by surrounding material, the \mbox{X-ray} observations are extremely efficient at identifying AGNs \citep[e.g.,][]{bran05}. In particular, the rest-frame 0.5--8.0~keV luminosity and the hard-to-soft \mbox{X-ray} band ratio are useful properties in distinguishing between AGNs and starbursts. We therefore use the  \emph{Chandra} \mbox{X-ray} source catalogs of the \mbox{CDF-S}, \mbox{E-CDF-S}, and EGS to calculate the X-ray properties of our sources.

For the \mbox{E-CDF-S} field, we use the 2~Ms \mbox{CDF-S} \mbox{X-ray} source catalog of \citet{luo08} and the 250~ks \mbox{E-CDF-S} \mbox{X-ray} source catalog of \citet{lehm05}. For the $\approx 200$~ks EGS, we use the \mbox{X-ray} source catalog of \citet{lair09}. Since a number of differences exist between the EGS catalog and the other two catalogs in the band definitions used to measure counts and fluxes (e.g., the full band is defined as 0.5--7.0~keV in the EGS catalog and 0.5--8.0~keV in the \mbox{CDF-S} and \mbox{E-CDF-S} catalogs), we used a simple power-law model to convert counts and fluxes to a uniform system. For these conversions, we first derive the effective power-law index from the band ratio given in the EGS catalog following the method used in \S3.3 of \citet{luo08}. Briefly, we find the power-law model (including an assumed Galactic column density) that reproduces the observed band ratio. For sources with a low number of counts \citep[$\lesssim 30$ counts total; for details, see][]{luo08}, we adopt $\Gamma=1.4$, a value representative of faint sources. We then use the effective power-law index to convert counts measured in the EGS bandpass to the corresponding \mbox{E-CDF-S} bandpass. X-ray luminosities and band ratios (the ratio of flux in the 2.0--8.0~keV band to that in the 0.5--2.0~keV band) were then calculated directly from the catalog fluxes.

\subsection{AGN Identification}\label{S:AGN_selection}
To identify AGNs among the \mbox{X-ray} sources, we follow the identification criteria used by \citet{baue04}, which we outline briefly here. AGNs were first identified based on their intrinsic, rest-frame 0.5--8.0~keV luminosities. An estimate of the intrinsic absorption is needed to derive this luminosity. By assuming that the AGN \mbox{X-ray} spectra are well represented by an intrinsic power-law with a photon index of 1.8, we can use the band ratio (the ratio of counts in the 2--8~keV band to the 0.5--2~keV band) to derive a basic estimate of the intrinsic $N_{\rm H}$ (see \S \ref{S:corrections} for details of the fitting procedure). Sources with rest-frame $L_{0.5-8.0\rm{~keV}} \gtrsim 3 \times 10^{42}$~erg~s$^{-1}$ are likely to be AGNs, since starbursts generally have luminosities below $L_{0.5-8.0\rm{~keV}} \lesssim 10^{42}$~erg~s$^{-1}$. However, to ensure that luminous starbursts are not misclassified as AGNs, we calculated the predicted 2--10~keV luminosities from star formation from the relations of \citet{pers07}. Persic et al.\ find that the scaling relation for ULIRGs is different than that of lower-SFR objects. Therefore, we use the following relation from \citet{pers07} to estimate the 2--10~keV luminosity due to star formation for systems with ${\rm SFR} \lesssim 100$~$M_{\odot}$~yr$^{-1}$:
\begin{equation}
L(2-10 \mbox{ keV})=3.8\times 10^{39} \frac{\rm SFR}{M_{\odot}\mbox{ yr$^{-1}$}} \mbox{ erg s$^{-1}$ }
\end{equation}
For systems with ${\rm SFR}\gtrsim 100$~$M_{\odot}$~yr$^{-1}$, \citet{pers07} found a somewhat different scaling was preferred:
\begin{equation}
L(2-10 \mbox{ keV})=0.75\times 10^{39} \frac{\rm SFR}{M_{\odot}\mbox{ yr$^{-1}$}} \mbox{ erg s$^{-1}$ }
\end{equation}
For the purposes of this calculation, we determined the star formation rate following \S\ref{S:SFRs} by assuming, conservatively, that the entire observed 70~\um\ flux is due to star formation. The predicted rest-frame 2--10~keV luminosity was converted to an observed-frame 0.5--8~keV flux assuming a $\Gamma=2$ power-law spectrum and the source redshift. We then classified as AGNs all sources with both $L_{0.5-8.0\rm{~keV}} \gtrsim 3 \times 10^{42}$~erg~s$^{-1}$ and with an observed luminosity $> 3$ times that predicted from the star formation rate.

Next, sources were classified by the hard-to-soft \mbox{X-ray} band ratio: sources with band ratios above 0.8 (corresponding to effective photon indices $\Gamma \lesssim 1$) were classified as AGN, as starbursts almost always have softer \mbox{X-ray} spectra (with $\Gamma \gtrsim 1$). Lastly, in addition to these purely \mbox{X-ray}--based criteria, we also use the \mbox{X-ray}--to--optical flux ratio as a further discriminator of AGN activity. We classify sources with $f_{0.5-8.0\rm{~keV}}/f_{R} >0.1$ as AGNs, where $f_R$ is the $R$-band flux. Using these criteria, we identified AGNs in 108 ($\approx 10$\%) of the 1022 70~\um\ sources for the combined \mbox{E-CDF-S} and EGS fields. The majority of identified AGNs meet more than one selection criterion. In particular, of 108 AGNs, 12 were identified uniquely using  $L_{0.5-8.0\rm{~keV}} \gtrsim 3 \times 10^{42}$~erg~s$^{-1}$ (which identified 94 AGNs in total), 9 using $\Gamma \lesssim 1$ (which identified 47 AGNs in total), and 7 using $f_{0.5-8.0\rm{~keV}}/f_{R} >0.1$ (which identified 83 AGNs in total). Therefore, we expect that few, if any, non-AGNs are misidentified as AGNs in our sample.

The remaining 50 non-AGN \mbox{X-ray} sources have properties consistent with starbursts and are considered to be such in the following analysis. The fraction of \mbox{X-ray} sources identified as starbursts in our sample ($\approx 30$\%) is higher than typical starburst fractions determined for flux-limited \mbox{X-ray} samples \citep[e.g.,][find $\approx 10$--20\% of the \mbox{CDF-S} \mbox{X-ray} sources are starbursts]{baue04}. However, these samples were selected differently from our mid-infrared-selected sample, which naturally includes a high fraction of luminous starbursts and should therefore be expected to have a higher fraction of \mbox{X-ray}-detected starbursts than would a comparable \mbox{X-ray} selected sample.

\subsubsection{Highly Obscured AGNs}\label{S:obscured_AGNs}
The selection criteria described above will find the bulk of the unobscured and moderately obscured AGNs to the \mbox{X-ray} survey flux limits. However, the most highly obscured or low-luminosity AGNs will be missed.  For example, \citet{dadd07II} and \citet{fior09}, among others, recently found tentative evidence for a population of highly obscured AGNs (potentially $N_{\rm H} \sim 10^{24.5}$~cm$^{-2}$) that lack significant \mbox{X-ray} emission. While the exact contribution from this highly obscured population to the total number of AGNs is unclear \citep[see e.g.,][]{alex08_ct,donl08}, they may represent a numerically important AGN population. Unfortunately, the reliable identification of such AGNs is difficult even in nearby sources and is beyond the scope of this work \citep[for a comprehensive analysis and review of infrared selection of AGNs, see][]{donl08}. We instead attempt to account for the effect that these missing AGNs have on the AGN fraction by adopting the intrinsic distribution of column densities determined by \citet{tozz06} and comparing it to the observed distribution of the AGNs identified in our 70~\um\ sample. A detailed discussion of our method is given in \S\ref{S:calc_AGN_fraction}.

\section{Derivation of AGN Properties and Star Formation Rates}\label{S:AGN_contrib}
In this section, we describe how we derived various AGN properties, such as the intrinsic column density and bolometric luminosity, that are required for an accurate determination of the AGN fraction. We also describe how we estimated the AGN contribution to the observed mid-infrared emission, which is required for deriving SFRs for these sources. 

\subsection{AGN Bolometric Corrections}\label{S:corrections}
The bolometric luminosity is commonly inferred from the mean, intrinsic energy distribution of a sample of representative AGNs \citep[e.g.,][]{elvi94}. The bolometric correction, which essentially gives the fraction of the bolometric luminosity emerging in a given band, can then be determined. Due to the relative ease with which AGN \mbox{X-ray} emission may be cleanly measured, the \mbox{X-ray} bolometric correction is commonly used  \citep[e.g.,][]{hopk07}. However, a number of complications exist with this method. First, knowledge of the intrinsic absorption is needed. For bright \mbox{X-ray} sources, the absorbing column density can be obtained directly through spectral fitting. For sources with few counts, the band ratio can provide a basic estimate of the absorption \citep[e.g.,][]{alex05}, but some uncertainty will remain. Second, the bolometric correction is known to vary with AGN luminosity, due to the luminosity dependence of the power-law slope between 2500~\AA\ and 2~keV (denoted $\alpha_{\rm OX}$), although this dependence is well quantified \citep[e.g.,][]{stef06}. Third, the intrinsic spread in AGN SEDs, possibly due in part to variability, is known to result in an uncertainty of a factor of several in the bolometric correction \citep{hopk07,vasu07}. 

For the purposes of this study, we use the luminosity-dependent \mbox{X-ray} bolometric corrections of \citet{hopk07}, determined for unobscured (type-1) quasars, which transform the intrinsic 2--10~keV luminosity to the bolometric luminosity and account for the luminosity dependence of the SED on $\alpha_{\rm OX}$, the power-law slope between 2500~\AA\ and 2~keV. To use this correction, we need estimates of the intrinsic luminosity and hence of the obscuring column densities for the AGNs in our sample. Recently, \citet{tozz06} performed \mbox{X-ray} spectral fitting of the AGNs in the \mbox{CDF-S} to derive reliable estimates of the column densities. We use the values of the column density derived by \citet{tozz06} when possible. For sources not included in the Tozzi et al.\ sample (80 of 108 AGNs), we estimate the column density using the observed \mbox{X-ray} band ratio for each source as follows.  

First, we model the \mbox{X-ray} emission using an absorbed power-law model in {\sc xspec} with both intrinsic and Galactic absorption. In {\sc xspec}, the model is defined as \textit{wabs}$\times$\textit{zwabs}$\times$\textit{zpow}. The photon index of the power-law component was fixed to 1.8 \citep[e.g.,][]{turn97} and the redshift of the \textit{zwabs} and \textit{zpow} components was fixed to that of the source. We additionally fixed the Galactic column density to $N_{\rm H} = 8.8 \times 10^{19}$~cm$^{-2}$ for the \mbox{E-CDF-S} \citep{star92} and to $N_{\rm H} = 1.3 \times 10^{20}$~cm$^{-2}$ for the EGS \citep{dick90}. Next, for each source, we used this model to find the intrinsic column density that reproduces the observed band ratio. To account for the \textit{Chandra} response, we extracted aimpoint responses from the event files used to construct the catalogs and used the appropriate responses during the fitting. Band ratios were taken from the catalogs, except in the EGS, where they were calculated from the provided counts and adjusted to the aim-point values using the source and aim-point effective exposure values. Approximately 10\% of the \mbox{X-ray} AGNs were not detected in the soft band, implying a very hard spectrum and resulting in a lower limit to the band ratio. For these sources, we use the column density that corresponds to the lower limit. 

\begin{figure}
\plotone{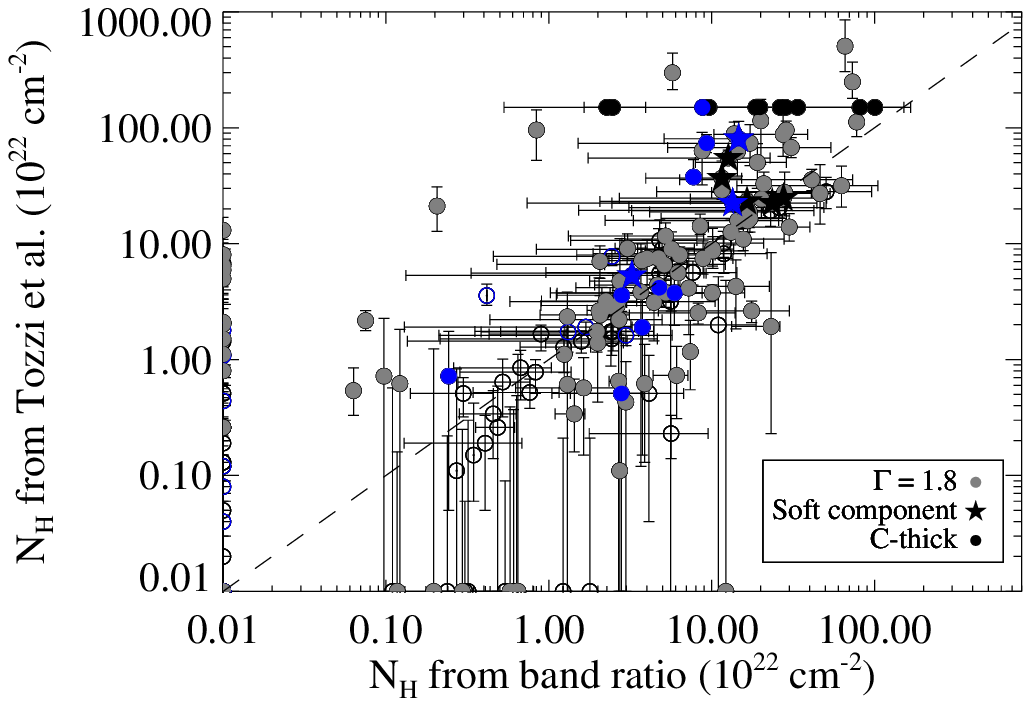}
\caption{Comparison of $N_{\rm H}$ values from \citet{tozz06} and from our analysis for \mbox{X-ray}-selected AGNs in the CDF-S. The source classification from \citet{tozz06} is indicated by symbol type, and 70~\um\ sources are indicated in blue. Sources with $N_{\rm H}<10^{20}$~cm$^{-2}$ are set to $N_{\rm H}=10^{20}$~cm$^{-2}$ for plotting purposes. A total of 11 sources have this value for both $N_{\rm H}$ estimates and appear as a single point in the lower left-hand corner. \label{F:tozzi}}
\end{figure}

Inferred column densities for the AGNs in our infrared-selected sample vary from $N_{\rm H}<10^{19}$~cm$^{-2}$ to $N_{\rm H}\approx 10^{24}$~cm$^{-2}$. In general, our values agree reasonably well with those derived by \citet{tozz06} for a sample of 194 \mbox{X-ray}-identified AGNs in the \mbox{CDF-S} with detections in both the hard and soft bands ($\sigma \approx 1$~dex; see Figure~\ref{F:tozzi}). Some of the scatter is due to differences between the redshifts used by Tozzi et al.\ and those used by us (which have been supplemented with new spectroscopic redshifts and the high-quality photometric redshifts discussed in \S\ref{S:redshifts}). For 53 ($\approx 25$\%) of the 194 sources, our redshifts differ by more than 20\% from those used by Tozzi et al.  Although the agreement between our values of the column density and those of Tozzi et al.\ is good overall, at values of $N_{\rm H}\gtrsim 3\times10^{23}$~cm$^{-2}$, our estimates of the column density appear to be systematically low. A number of these sources were identified by Tozzi et al.\ as Compton-thick. Unfortunately, due to the typically faint \mbox{X-ray} fluxes for such sources, their reliable identification is difficult \citep[e.g.,][]{tozz06} and beyond the scope of this paper.

Since our sources were chosen to be luminous 70~\um\ emitters, it is possible that the \mbox{X-ray} flux includes a contribution from star formation. Such a contribution could result in derived values of the column density that are systematically low. To determine if this effect is likely to be significant in the objects in our sample, we follow the procedure used in \S\ref{S:AGN_selection} to estimate the expected contribution from star formation using the relations of \citet{pers07}. As before, we estimated upper limits on the star formation rates by assuming that the entire observed 70~\um\ flux is due to star formation. We find that the predicted contribution from star formation to the observed 2--8~keV luminosity is $\lesssim 10$\% in all cases, with a mean value of $\approx 0.8$\%. Therefore, in our AGN-selected sample, the AGN emission likely dominates in the \mbox{X-ray} band, and emission related to star formation is not expected to have a significant effect on the derived column densities. This conclusion is supported by Figure~\ref{F:tozzi}, in which the column densities of AGNs in 70~\um\ sources (shown in blue) do not differ systematically from those of the whole population, with the exception of the lowest column densities ($N_{\rm H} \lesssim 10^{21}$~cm$^{-2}$), where our values of the column density do appear to be systematically low. However, the corrections required to correct the observed fluxes for these low column densities are modest and have little effect on our derived bolometric luminosities.

\begin{figure}
\plotone{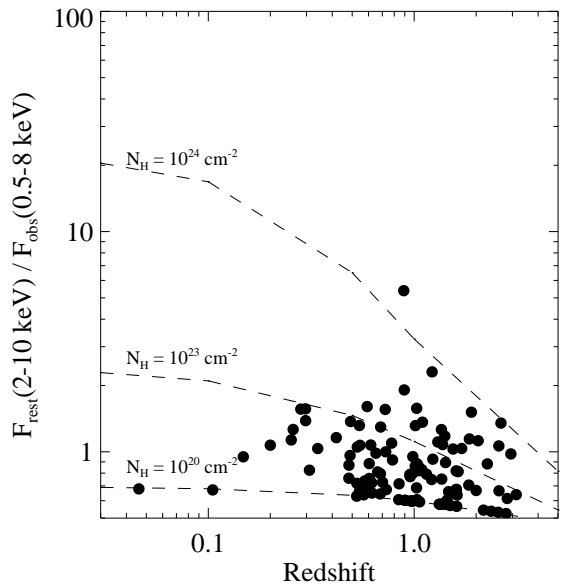}
\caption{The correction from observed 0.5--8~keV flux to rest-frame 2--10~keV flux as a function of redshift for the \mbox{X-ray}-selected AGNs. \label{F:flux_corrections}}
\end{figure}

The resulting column densities and the absorbed power-law model described above were used to calculate corrections to transform the observed-frame 0.5--8~keV fluxes to unabsorbed, rest-frame fluxes. To illustrate the range of corrections that we find, we plot the corrections required to transform from observed 0.5--8~keV fluxes to rest-frame 2--10~keV fluxes as a function of redshift in Figure \ref{F:flux_corrections} \citep[cf.][]{alex08_bh}. The corrections generally range from $\approx 0.5$--2, but one source with a high column density requires a correction of $\approx 5$--6. We then estimate the bolometric correction required to scale the rest-frame 2--10~keV luminosity to a bolometric luminosity from the models of \citet{hopk07}.

Lastly, we can obtain basic fiducial estimates of the black-hole accretion rate from the bolometric luminosity by assuming an efficiency for the conversion of the rest mass of the accreting material to luminosity. We adopt an efficiency of $\epsilon=0.1$, typical of AGNs accreting at rates $\sim 10$\% or more of the Eddington rate \citep[e.g.,][]{marc04}. The accretion rate is then $\dot{M}_{\rm acc} = L_{\rm bol}/(\epsilon c^2),$ where $c$ is the speed of light.

\subsection{Mid-Infrared AGN SEDs}\label{S:agn_seds}
Driven by studies of the infrared background, a great deal of work has gone into constructing model infrared AGN SEDs from the observed SEDs of large samples of AGNs. We consider three recent model SEDs to estimate the AGN contribution to the mid-infrared emission: the type-1 and type-2 AGN models of \citet{silv04}, the type-1 AGN models of \citet{hopk07}, and a mean SED of AGNs from a flux-limited \textit{Swift} BAT survey \citep{tuel08,wint09}.

\citet{silv04} constructed type-1 and type-2 AGN infrared SEDs using a sample of 33 Seyfert galaxies and 11 quasars with available nuclear mid-infrared and \mbox{X-ray} fluxes. They constructed intrinsic SEDs by interpolating the observed SEDs (up to rest-frame $\lambda \approx 20$~\um). Beyond $\lambda \approx 20$~\um, they extrapolated from the observed SEDs using the radiative-transfer models of \citet{gran94} for a number of different absorbing column densities. In a more recent study, \citet{hopk07} construct a model SED for type-1 quasars using a number of components at different wavelengths, including a mean optical spectrum and a power-law \mbox{X-ray} spectrum. In the infrared, they adopt the mean spectrum from \citet{rich06}. Lastly, \citet{mull09} use a sample of 36 AGNs detected with \textit{Spitzer} and the \textit{Swift} BAT from the sample of \citet{wint09}, selected to have no strong indication from PAH features of a significant contribution from star formation to the mid-infrared emission. These \mbox{X-ray} selected AGNs should be fairly representative of the AGNs in our sample. Mullaney et al.\ have constructed an average mid-infrared SED from these sources using a combination of \textit{Spitzer} IRS spectra and IRAS photometry (which provide coverage at wavelengths beyond those covered by the IRS spectra). These three studies provide some of the best determinations of the mid-infrared SEDs of AGNs currently available.

To normalize the AGN SEDs to the observed \mbox{X-ray} luminosity, we assume that the mid-infrared SED scales linearly with the bolometric AGN luminosity over the luminosity range of our sample \citep[i.e., that the AGN luminosity is the principle determinant of the infrared luminosity; see, e.g.,][]{haas03}. This approximation holds well for the luminosity-dependent AGN SEDs of \citet{hopk07}, which include effects such as the dependence of $\alpha_{\rm OX}$ on the AGN luminosity \citep[e.g.,][]{stef06}. The bolometric luminosity was derived following \S\ref{S:corrections} and scales with the intrinsic \mbox{X-ray} luminosity approximately as $L_{\rm bol}^{\rm AGN} \propto L_{\rm 2-10~keV}^{1.39}$ over the luminosity range of our sample. The models of \citet{silv04} and the average BAT AGN SED that we use are luminosity independent (i.e., their shapes do not change as a function of AGN luminosity). For consistency with the Hopkins et al.\ models, we scaled the type-1 AGN Silva et al.\ models to have the same 2500~\AA\ luminosity as the (type-1 AGN) Hopkins et al.\ models at a given bolometric luminosity. This same scaling was also used when scaling the type-2 AGN Silva et al.\ models. For the BAT AGN SED, the scaling was set such that the average $L_{\rm IR}$ of the BAT sample \citep[calculated from the observed fluxes using the relation of][]{sand96} is recovered correctly from the bolometric luminosity corresponding to the average $L_{\rm X}$ of the BAT sample. 

\begin{figure*}
\plotone{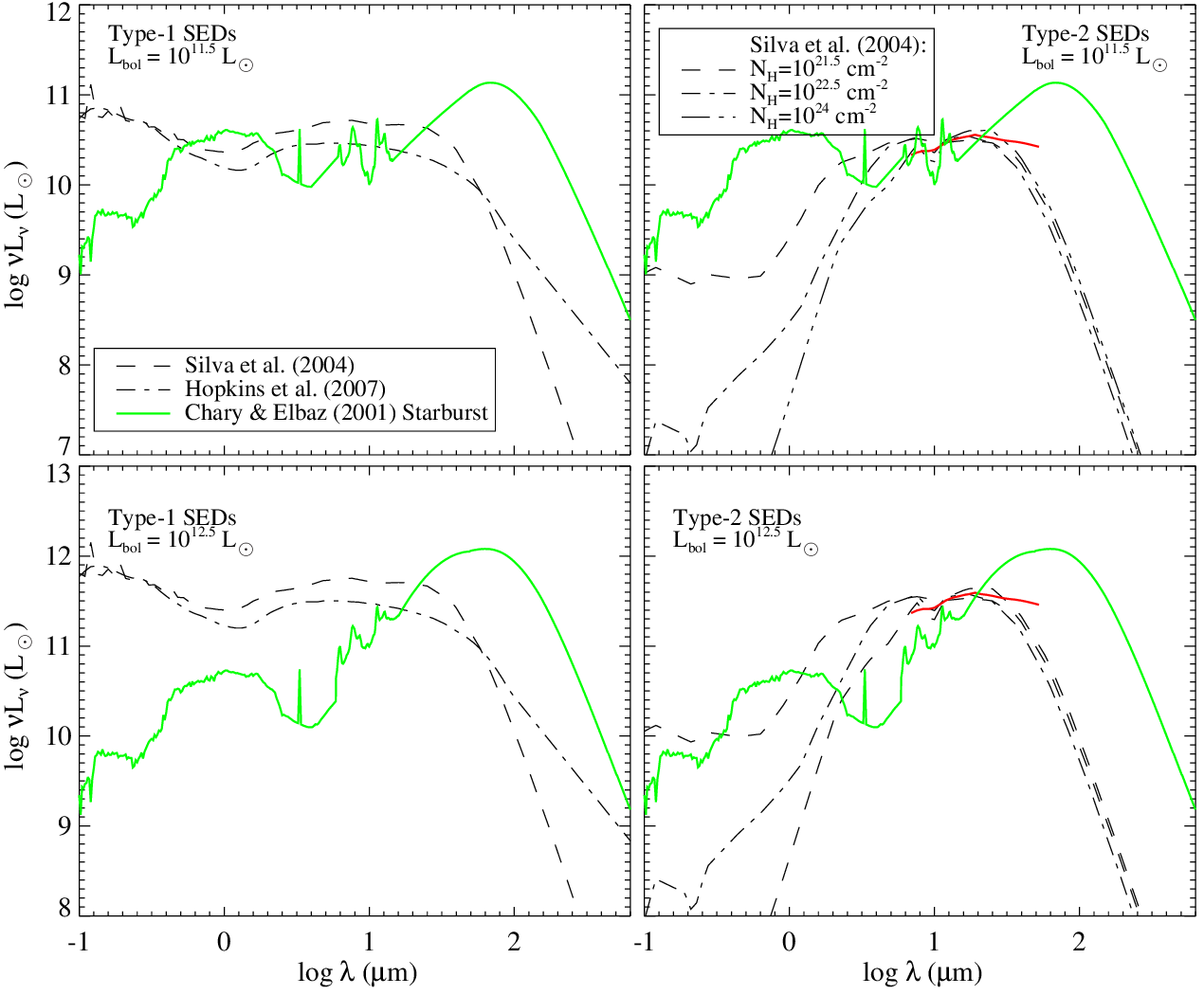}
\caption{Comparison of various AGN SED models (denoted by dashed or dashed-dotted lines). The type-1 AGN models of \citet{silv04} and \citet{hopk07} are shown in the left panels, and type-2 AGN models of \citet{silv04} with range of intrinsic column densities are shown in the right panels. Also shown (in red) is the average SED from the BAT AGN sample (see \S\ref{S:agn_seds} for details). For comparison, the solid green line shows a starburst SED from \citet{char01} with a SFR chosen such that the 24~\um\ luminosity approximately matches that of the AGN SEDs. \label{F:AGN_SEDs}}
\end{figure*}

Figure \ref{F:AGN_SEDs} compares the three AGN models (scaled as described above) for AGN bolometric luminosities of $\log (L_{\rm bol}/L_{\odot}) = 11.5 $ and $\log (L_{\rm bol}/L_{\odot}) = 12.5 $.  The left two panels show type-1 AGN SEDs and a starburst SED \citep[from][]{char01} chosen to have roughly the same rest-frame 24~\um\ luminosity as the AGN SEDs. The two AGN SEDs agree to within a factor of 2--3 over the infrared region (the Silva et al.\ models predict higher infrared flux out to $\approx 70$~\um). As the difference between the two type-1 AGN models is smaller than the expected systematic errors, we adopt the more recent model of Hopkins et al.\ for subsequent analysis of the type-1 AGNs. The starburst SED, while having approximately the same rest-frame 24~\um\ luminosity as the AGN SEDs, clearly dominates at longer wavelengths, due to the lower temperature of dust the reprocesses emission from young stars. 

The right panels of Figure \ref{F:AGN_SEDs} show the average BAT AGN SED and the type-2 AGN SEDs of \citet{silv04} for a variety of $N_{\rm H}$ values, again with starburst SEDs chosen to have approximately the same 24~\um\ luminosity as those of the AGNs overlaid. At higher values of $N_{\rm H}$, the Silva et al.\ type-2 AGN SEDs show heavy extinction at wavelengths below $\sim 10$~\um\ compared to the type-1 AGN SEDs shown in the left panels. In general, however, between $\sim 20$--70~\um\ (the approximate range probed by our observed-frame 70~\um\ data) the Silva et al.\ SEDs are very similar, both among the type-2 AGN models of different $N_{\rm H}$ and when compared to the type-1 AGN model. The average BAT AGN SED, however, has higher luminosity at wavelengths beyond $\sim 40$~\um\ (by up to a factor of $\sim 3$ over the probed wavelength range) compared to the Silva et al.\ models. This difference will result in larger predicted observed-frame 70~\um\ fluxes for type-2 AGNs with $z \lesssim 0.75$ when the BAT AGN SED is used. For sources at higher redshift (which include most of the high-luminosity sources), the predicted 70~\um\ fluxes from the two models will agree closely. As this difference will primarily affect only the lower-luminosity sources (as low-redshift sources tend to have lower luminosities), which generally have low predicted AGN contributions, our results are not significantly changed from those obtained using the Silva et al.\ models. Therefore, we adopt the Silva et al.\ models for the type-2 AGNs models used in the next section. 

\subsection{Predicted AGN contribution to the Mid-Infrared Flux}\label{S:pred_AGN_fluxes}
\begin{figure}
\plotone{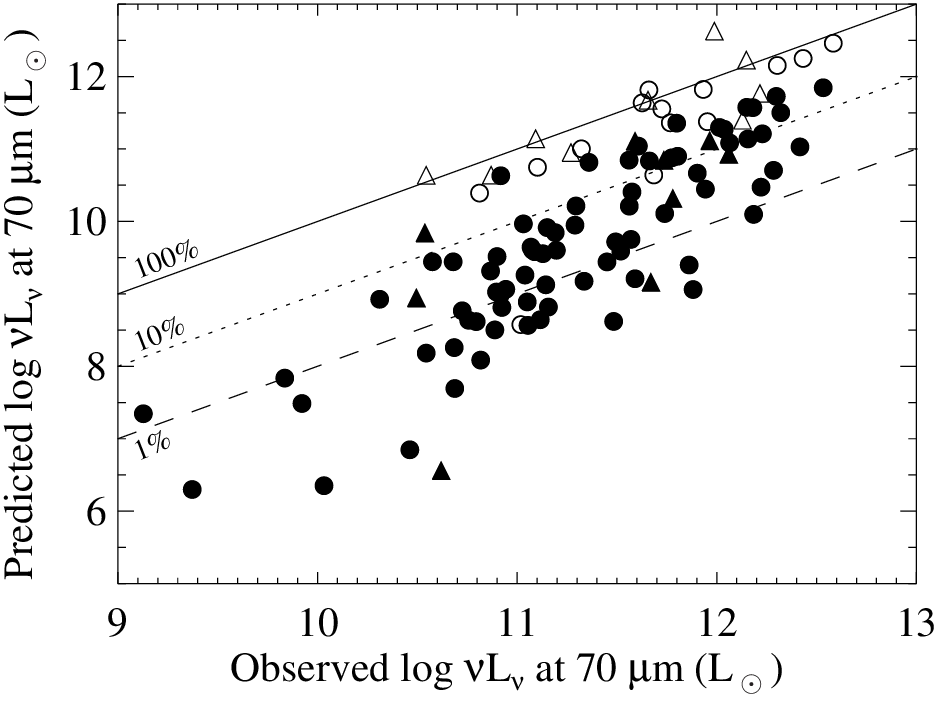}
\caption{Comparison of the AGN's predicted observed-frame 70~\um\ luminosity to the observed 70~\um\ luminosity using the models of \citet{silv04} for non-type-1 AGNs (\textit{circles}) and the models of \citet{hopk07} for type-1 AGNs (\textit{triangles}). Open symbols denote systems that have a net ${\rm S/N}< 3$ after subtracting the estimated contribution from the AGN (see text for details). The solid line denotes equality, and the dashed and dotted lines indicate an AGN contribution to the observed 70~\um\ luminosity of 1\% and 10\%, respectively. \label{F:70um_comparison}}
\end{figure}

We can use the model AGN SEDs to predict the AGN contribution to the observed 70~\um\ flux. Before doing so, we divided the AGN sample into subsamples based on the presence of type-1 AGN optical characteristics in either the \citet{baue04} catalog, the COMBO-17 catalog, or the DEEP2 catalog. In these catalogs, non-type-1 AGNs are simply AGN that are not clearly type-1 AGNs, so the non-type-1 AGNs are likely to include some type-1 AGNs with more subtle characteristics. In total, 18 ($\approx 17$\%) of the 108 AGNs in our sample were identified as type-1 AGNs. We used type-1 AGN models to predict the mid-infrared fluxes of the type-1 AGNs and type-2 AGN models with the appropriate intrinsic column density to predict the mid-infrared fluxes of the non-type-1 AGNs. We then used the intrinsic, unabsorbed 2--10~keV flux derived earlier (see \S\ref{S:corrections}) to normalize the models (effectively a bolometric correction) and used linear interpolation (in log space) of the model SEDs to derive the observed-frame 70~\um\ flux or luminosity. In Figure \ref{F:70um_comparison}, we compare the observed 70~\um\ luminosity to the one predicted by the models. In general, the predicted 70~\um\ luminosity is much lower than that observed for the majority of AGNs in our sample. 

\begin{figure}
\plotone{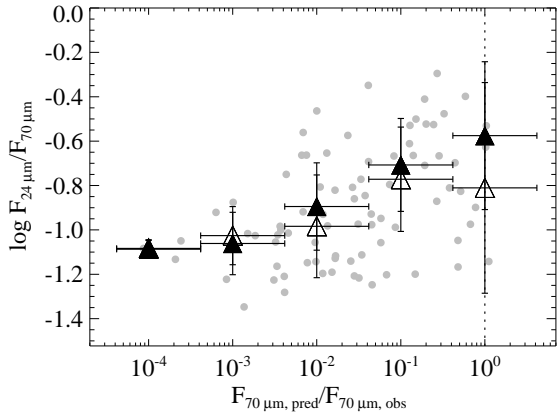}
\caption{The observed mid-infrared color versus the ratio of (observed-frame) predicted-to-observed 70~\um\ flux for the AGNs in the sample with $z<1.5$. The filled points show the mean values (calculated using the non-logarithmic values of the colors) of subsamples of objects in bins with widths indicated by the horizontal error bars. The open points show the median values for the same bins. Vertical error bars indicate the standard deviation of the colors in each bin.  \label{F:ircolor_vs_ratio}}
\end{figure}

Critical to this comparison is the normalization of the SED, which depends on an accurate estimate of the intrinsic \mbox{X-ray} flux. As discussed in \S\ref{S:corrections}, our sample of \mbox{X-ray}-detected AGNs does not show evidence from \mbox{X-ray} band ratios of being highly extincted (the majority of inferred column densities are $\lesssim 2 \times 10^{23}$~cm$^{-2}$). Therefore, the corrections required to convert observed \mbox{X-ray} fluxes to unabsorbed, intrinsic fluxes are typically modest and should not be subject to large uncertainties. Indeed, it is clear from Figure \ref{F:70um_comparison} that the predicted (observed-frame) AGN 70~\um\ luminosity (derived from the redshifted AGN model) is consistent with the observed one in all but one system. For this one system (in which the predicted luminosity exceeds the observed one by a factor of $\sim 3$), \mbox{X-ray} variability may account for the discrepancy, if this AGN was observed in a state of higher-than-average \mbox{X-ray} luminosity (conversely, variability may also lead to underestimates of the AGN luminosity in some sources observed in lower-than-average states). We emphasize that, because our results depend only on properties averaged over many systems, they should not be strongly affected by such variability. To illustrate this point, in Figure~\ref{F:ircolor_vs_ratio} we plot our predicted average fractional 70~\um\ contributions from the AGN to the observed flux, binned on 1-dex bins, against the observed mid-infrared color for the AGNs in our sample with $z<1.5$. The mid-infrared color has been found to be a rough indicator of the AGN contribution, with warmer colors indicating higher AGN contributions, at least out to $z\sim 1.5$; beyond this redshift, the color is less reliable \citep[e.g.,][]{mull09}. There is a clear trend between the two indicators: systems with high predicted AGN contributions to the observed 70~\um\ flux tend to have warmer colors than systems with low predicted AGN contributions. Therefore, on average, it appears that our method produces estimates of the AGN contribution to the mid-infrared flux that are generally consistent with dust temperatures indicated by the mid-infrared color. Further comparisons to other estimates of the relative AGN contribution, such as those from spectra decomposition of mid-infrared spectra, will provide useful tests of systematic errors in our method. However, such analyses are beyond the scope of this work.


Additionally, as the AGNs can contribute significantly to the observed mid-infrared emission, sources of a given SFR and redshift that host AGNs will be detected more readily than those that lack AGNs. To avoid biasing our SFR-selected sample toward systems with AGNs, we constructed an unbiased sample (henceforth known as the ``SFR sample'') by eliminating sources in which the net 70~\um\ flux (after subtracting the AGN's contribution) results in a signal-to-noise ratio that falls below our adopted limit ($S/N=3$; see \S\ref{S:sample}). We found that 23 ($20$\%) of 108 AGNs fell below this limit in the combined \mbox{E-CDF-S} and EGS samples (shown as open symbols in Figure~\ref{F:70um_comparison}). The SFR sample is used only to study the AGN activity as a function of SFR. When we examine the AGN activity as a function of other properties (e.g., mid-infrared color), the full sample of 108 AGNs is used. We note that most of the AGNs that were eliminated have $\log \left( F_{\rm 24 \mu m}/F_{\rm 70 \mu m} \right)>-0.7$ and are predicted to contribute a large fraction ($\gtrsim 50$\%) of the 70~\um\ flux. In the SFR sample of 85 AGNs (of 1022 70~\um\ sources in total), only two AGNs have an estimated contribution to the observed 70~\um\ flux of more than 50\%. Therefore, this cut eliminates most of the sources for which the determination of the SFR is likely to be subject to large systematic uncertainties (i.e., those in which AGN-powered emission likely dominates at 70~\um).

Lastly, due to intrinsic differences of the SEDs of systems with the same SFR, some systems at a given SFR will have 70~\um\ fluxes that fall below our adopted flux limit, resulting in our sample being incomplete at the given SFR. We correct for this incompleteness by estimating the scatter about the relation used by \citet{char01} for local starbursts. We estimated a scatter of $\sim 0.5$~dex in the observed flux at 30--70~\um\ at all star formation rates, and we have used this value to estimate the likely number of missed starbursts for each detected one. We did this by generating, for each detected source, a normal distribution of fluxes around the observed 70um flux. We then calculate the fraction of sources that fall below the flux limit for the detected source under consideration and correct the total number of starbursts by the sum over all sources. We assume that these missed sources, since they are presumably cooler than the average starburst for a given LIR, do not host AGNs. Again, as in the rest of our analysis, we do not include any evolution in the starburst SEDs, as such evolution is currently poorly understood. This correction is generally small, and is significant only for those sources detected near the flux limit. Additionally, where the AGN fraction is high (such as at high SFRs), the effect of this correction is reduced (since the fraction of non-AGN hosts, which is being adjusted, is by definition lower).

\subsection{Star Formation Rates}\label{S:SFRs}
As discussed in \S1, mid-to-far-infrared observations sample a significant fraction of the energy emitted by massive stars in dusty environments, with much of the remainder emerging at UV wavelengths. Therefore the total star formation rate may be estimated as the sum of the rate inferred from direct UV emission and the rate inferred from the reprocessed, infrared emission \citep[e.g.,][]{dadd07II}.

To trace the infrared emission associated with star formation, we use the observed-frame 70~\um\ luminosity. A number of recent studies \citep[e.g.,][]{shi07,tadh07,vega07} have found that, at a given star formation rate, the observed-frame 70~\um\ emission suffers from significantly less AGN contamination and spectral complexity than 24~\um\ or shorter-wavelength emission, particularly at higher redshifts ($z\gtrsim 1.5$). However, the contribution from AGNs to the 70~\um\ emission can still be significant. Therefore, for sources hosting an AGN, we estimated the AGN contribution to the 70~\um\ luminosity using empirical AGN SEDs (see \S\ref{S:AGN_contrib} for details). The net observed-frame 70~\um\ luminosity was then converted to a rest-frame, 8--1000~\um\ luminosity (denoted $L_{\rm IR}$) using the dusty starburst models of \citet{char01}, which are luminosity dependent, and the prescription of \citet{sand96}. The resulting $L_{\rm IR}$ was then converted to a SFR using the relation of \citet{kenn98}, which assumes a \citet{salp55} IMF, as follows:
\begin{equation}\label{E:kennsfr}
\frac{{\rm SFR}_{\rm IR}}{M_{\odot}\rm{~yr}^{-1}} = 1.73\times10^{-10}\left( \frac{L_{\rm IR}}{L_{\odot}} \right).
\end{equation}

We also investigated the use of other publicly available dusty starburst models and found that, for our sample and using the observed-frame 70~\um\ flux (which samples rest-frame wavelengths $\gtrsim 20$~\um), there is little practical difference ($\lesssim 50$\% in $L_{\rm IR}$) between the Chary \& Elbaz models that we adopt and those of \citet{dale02} or \citet{rieke09}.\footnote{If, however, we were to use the observed-frame 24~\um\ flux, which would sample rest-frame wavelengths of $\gtrsim 5$~\um\ for our sample, the differences between models can become large (up to factors of $\gtrsim 10$ in $L_{\rm IR}$), as illustrated by \citet{sali09}.} We note that all of these models are derived from local samples of starburst galaxies and hence could differ systematically from the sources in our study (which are generally at redshifts of 0.5--1.5). However, \citet{elba02}, in a study of infrared-luminous galaxies, found good agreement out to $z\sim 1$ between the radio-derived SFRs and those derived from the infrared emission. At $z\sim 2$, \citet{dadd07I} also find reasonable agreement between various indicators of star formation rate, including the infrared \citep[we note, however, that  sub-millimeter galaxies, at $z \gtrsim 2$, likely have lower dust temperatures for a given SFR than local ULIRGs, e.g.,][]{pope07,copp08}.  Since the choice of model makes little difference for the values of $L_{\rm IR}$ that we derive and a clear consensus as to the most appropriate model for high-redshift, high-luminosity sources has yet to emerge, we adopt the models of Chary \& Elbaz for all further analysis. 

As stated above, an additional important component of the bolometric emission from star formation emerges in the UV. This ``direct'' emission must be included when deriving the total SFR. To this end, the rest-frame UV luminosity was estimated using the rest-frame \mbox{$B$-,} \mbox{$V$-,} and $R$-band fluxes derived from the observed SED by fitting galaxy templates using ZEBRA or by simple linear interpolation (when possible). The UV conversions from two recent studies, \citet{dadd04} and \citet{bell05}, were used to transform the UV luminosity to a SFR. The Daddi et al.\ relation uses the 1500~\AA\ rest-frame luminosity to calculate the UV SFR as ${\rm SFR}_{1500}/ (M_{\odot}\mbox{~yr}^{-1}) = 1.13\times 10^{-28} (L_{1500}/L_{\odot})$. Bell et al.\ use the 2800~\AA\ luminosity to estimate the UV SFR as ${\rm SFR}_{2800}/ (M_{\odot}\mbox{~yr}^{-1})=8.99\times 10^{-29} (L_{2800}/L_{\odot})$. At $z \lesssim 1.5$, the 1500~\AA\ rest-frame emission is sampled only by GALEX observations. Therefore, to avoid large extrapolations for sources at $z \lesssim 1.5$, we use the Daddi et al.\ relation only for those sources that have GALEX detections ($\approx 40$\%).  The Daddi et al.\ and Bell et al.\ estimates, which typically agree to within a factor of a few, were then averaged to obtain the UV SFR, ${\rm SFR}_{\rm UV}$. The total SFR is then calculated as ${\rm SFR}_{\rm tot}={\rm SFR}_{\rm IR} +{\rm SFR}_{\rm UV}$. It should be noted that no correction is applied to account for extinction in the UV, as emission that is absorbed by dust will be reprocessed and is therefore included in the infrared-derived SFR. Since emission from the AGN may dominate at UV wavelengths, we assumed that the fraction of emission emerging in the UV from star formation for the AGN sources is the same as that for the non-AGN sample on average (i.e., the extinction in the UV is similar). For the sample as a whole, the UV SFRs generally represent $< 50$\% of the total SFR. Figure~\ref{F:z_dist}c shows the distribution of SFRs. It is clear from this figure that the sample is approximately complete to $z=1.0$ ($z=2.0$) for sources with ${\rm SFR} \gtrsim 100$~$M_{\odot}$~yr$^{-1}$ (${\rm SFR} \gtrsim 600$~$M_{\odot}$~yr$^{-1}$).

\section{Calculation of the AGN Fraction}\label{S:calc_AGN_fraction}
The AGN fraction is defined as the number of AGNs above a given intrinsic \mbox{X-ray} luminosity divided by the total number of sources in which an AGN was detected or could have been detected, given the sensitivity limits of the \mbox{X-ray} observations \citep[e.g.,][]{lehm08}. The cumulative AGN fraction may then be calculated following \citet{silv08} so that the contribution of each AGN to the total fraction is included:
\begin{equation}\label{E:agn_fraction}
f_{\rm AGN}=\sum_{i=1}^{N_{\rm AGN}} \frac{1}{N_{{\rm gal},i}}
\end{equation}
In this equation, $N_{\rm AGN}$ is the total number of AGNs in the sample and $N_{{\rm gal},i}$ is the number of galaxies in which the $i$-th AGN could have been detected. We further restrict $N_{{\rm gal},i}$ to include only those sources that lie in regions of sensitivity great enough to detect (at a $S/N > 3$) the 70~\um\ flux of the $i$-th AGN, thereby imposing a flux limit (as opposed to a $S/N$ limit) on the sources that contribute to each AGN's contribution to the total fraction. The error in the AGN fraction is calculated \citep[again following][]{silv08} as:
\begin{equation}\label{E:agn_fraction_err}
\sigma_{f}^{2} \approx \left( \sum_{i=1}^{N_{\rm AGN}} \frac{1}{N^2_{{\rm gal},i}} \right) + \sigma_{\rm phys}^2 + \sigma_{N_{\rm H}}^2,
\end{equation}
where $\sigma_{\rm phys}$ is the contribution to the error from uncertainties in the physical properties used to define the bins (the SFR, the rest-frame mid-infrared luminosity, and the mid-infrared color) and $\sigma_{N_{\rm H}}$ is the uncertainty resulting from the probabilistic treatment of the intrinsic $N_{\rm H}$ distribution (discussed in detail later in this section). The $\sigma_{\rm phys}$ term is estimated using a Monte Carlo technique as follows. For each source, we drew random values of the physical property from a normal distribution centered on the measured value of the physical property with a standard deviation given by the uncertainty in the property (e.g., for the SFR, we used errors derived from the reported errors in the 70~\um\ fluxes). We repeated this procedure 100 times, each time calculating a new fraction, and estimated $\sigma_{\rm phys}$ from the resulting distribution of $f_{\rm AGN}$.

Because the sensitivity of the \mbox{X-ray} data used in this study varies with position (by factors of $\gtrsim 10$), systematic errors will be induced in the AGN fraction if this variation is not accounted for when determining $N_{{\rm gal},i}$. To remove the effects of \mbox{\mbox{X-ray}} sensitivity variations, we include in $N_{{\rm gal},i}$ only those galaxies in which an AGN with luminosity $L_{{\rm X},i}$ could have been detected if present (i.e., only galaxies with limiting \mbox{X-ray} luminosities below $L_{{\rm X},i}$). To estimate the \mbox{X-ray} sensitivity limits, we used sensitivity maps generated separately for each field. Due to the dependence of the \textit{Chandra} point spread function and effective area on the off-axis angle, the \mbox{X-ray} sensitivity across a single field is a strong function of the position relative to the aim point of the observations. This variation can be estimated and, under the assumption of Poisson statistics \citep[e.g.,][]{luo08}, maps may be generated that give the sensitivity limit of the survey as a function of position. For the \mbox{CDF-S} and \mbox{E-CDF-S}, maps were generated as part of the catalog construction \citep[see][]{lehm05,luo08} in terms of the limiting flux that corresponds to the number of counts required for the secure detection of a source with a $\Gamma=1.4$ power-law spectrum. For the EGS, maps were provided directly in terms of limiting counts for a source with the same spectrum (see Laird et al.\ 2009). If a source lies in the \mbox{CDF-S} region (and therefore has both 2~Ms and 250~ks coverage), we adopted the lowest limiting flux from the \mbox{CDF-S} and \mbox{E-CDF-S} sensitivity maps at that position. This flux is generally that of the 2~Ms \mbox{CDF-S} except in some regions at large ($\gtrsim 8$\arcmin) angles from the average \mbox{CDF-S} aim point that lie near the \mbox{E-CDF-S} aim points. 

Next, we attempt to account for the distribution of intrinsic AGN column densities, which will affect both the overall AGN fraction (as high-column-density sources will tend to be missed in even the deepest \mbox{X-ray} surveys) and will result in field-to-field variations in the fraction due to different exposure times that probe different column densities at a given redshift. For example, at a given luminosity and redshift, AGNs with larger intrinsic column densities can be detected near the center of the \mbox{CDF-S} (where exposure times reach $\approx 2$~Ms) than in the shallower \mbox{E-CDF-S} (where typical exposure times are $\approx 250$~ks). We account for these effects by using an estimate of the intrinsic distribution of AGN column densities to effectively adjust the detection limits derived from the sensitivity maps. We use the intrinsic AGN $N_{\rm H}$ distribution determined for the \mbox{CDF-S} by \citet{tozz06}, who found that the distribution can be approximated by a log-normal distribution with $\left< \log N_{\rm H}/({\rm cm}^{-2}) \right> \approx 23.1$ and $\sigma \approx 1.1$. 

In deriving this distribution, Tozzi et al.\ assumed there is no strong dependence on the distribution with intrinsic luminosity or redshift. However, there is some evidence that the absorbed fraction is lower for higher luminosity AGNs \citep[e.g.,][]{ueda03,trei05,hasi08}. Given the current uncertainties in the detailed dependence of the $N_{\rm H}$ distribution on luminosity and redshift, we do not attempt to account for any such dependence but note that the Tozzi et al.\ distribution was derived using an \mbox{X-ray}-selected AGN sample that is similar to our AGN sample (and shares our \mbox{CDF-S} sources) and should therefore apply well to our sample on average. We slightly modified the log-normal distribution described above to match better the actual one found by \citet{tozz06} by maintaining a flat distribution from $N_{\rm H} \approx 10^{23}$~cm$^{-2}$ to $N_{\rm H} \approx 10^{24}$~cm$^{-2}$ and by including the $\approx 10$\% of objects at low values of $N_{\rm H}$ ($\lesssim 10^{20}$~cm$^{-2}$). We note that this distribution is also generally consistent with that adopted in other recent studies \citep[e.g.,][]{gill07,merl08} and in the \citet{hopk07} study from which we have derived the AGN bolometric corrections.

Using the sensitivity maps described above, we can now determine the number of 70~\um\ sources in which an AGN of a given luminosity, subject to the adopted $N_{\rm H}$ distribution, could have been detected. We first draw 1022 values of $N_{\rm H}$ from the above distribution and assign these values to each source. Then, for the $i$th AGN in our sample, we place hypothetical AGNs with intrinsic 0.5--8~keV luminosities equal to $L_{{\rm X},i}$ in all 70~\um\ sources. We then calculate the resulting absorbed, observed-frame 0.5--8~keV fluxes for each hypothetical AGN using the source redshifts, the assigned column densities, and the absorbed power-law AGN model described in \S\ref{S:corrections}. Although this model does not include the scattered, reflected, or line emission identified in many highly obscured AGNs \citep[e.g.,][]{mali03} and hence may underpredict the soft flux emerging from sources assigned high values of $N_{\rm H}$ (above $\sim 10^{24}$~cm$^{-2}$), we found no significant difference in our results when an empirically motivated AGN model \citep[constructed following][]{alex05b} was used. Therefore, for simplicity, we adopt the absorbed power-law model in our analysis.

We can now compare the predicted observed flux for each hypothetical AGN with the sensitivity limit at its position to determine whether or not such an AGN could have been detected if present. Since the \textit{Chandra} response is a strong function of energy, and AGNs with higher values of $N_{\rm H}$ have harder spectra, the resulting number of predicted 0.5--8~keV \textit{Chandra} counts (which determines whether or not a source will be detected) will differ for two sources with similar observed-frame 0.5-8~keV fluxes and redshifts but different values of $N_{\rm H}$. Therefore, for each 70~\um\ source, we calculate the number of detected counts expected from a given hypothetical AGN using the absorbed power-law AGN model (at the appropriate redshift), the appropriate \textit{Chandra} response, and the effective exposure time at the position of the source. We then compare the predicted number of counts to the limiting number of counts at the source position. If the predicted number of counts exceeds the limiting number, the source is included in the calculation of the AGN fraction.  We use the full-band (0.5--8~keV for the \mbox{CDF-S} and \mbox{E-CDF-S} and 0.5--7~keV for the EGS) limiting counts given by the sensitivity maps described above. For the \mbox{CDF-S} and \mbox{E-CDF-S}, for which the maps are given in terms of a limiting flux calculated assuming a $\Gamma=1.4$ power-law spectrum, we convert the flux to counts using the same spectrum and the effective exposure times at the positions of the sources. Finally, we repeat this entire procedure 100 times, and adopt the mean value of the fraction, $<f_{\rm AGN}>$, as the best estimate and set $\sigma_{N_{\rm H}}$ to the standard deviation of the fraction over the 100 runs.

\begin{figure*}
\plottwo{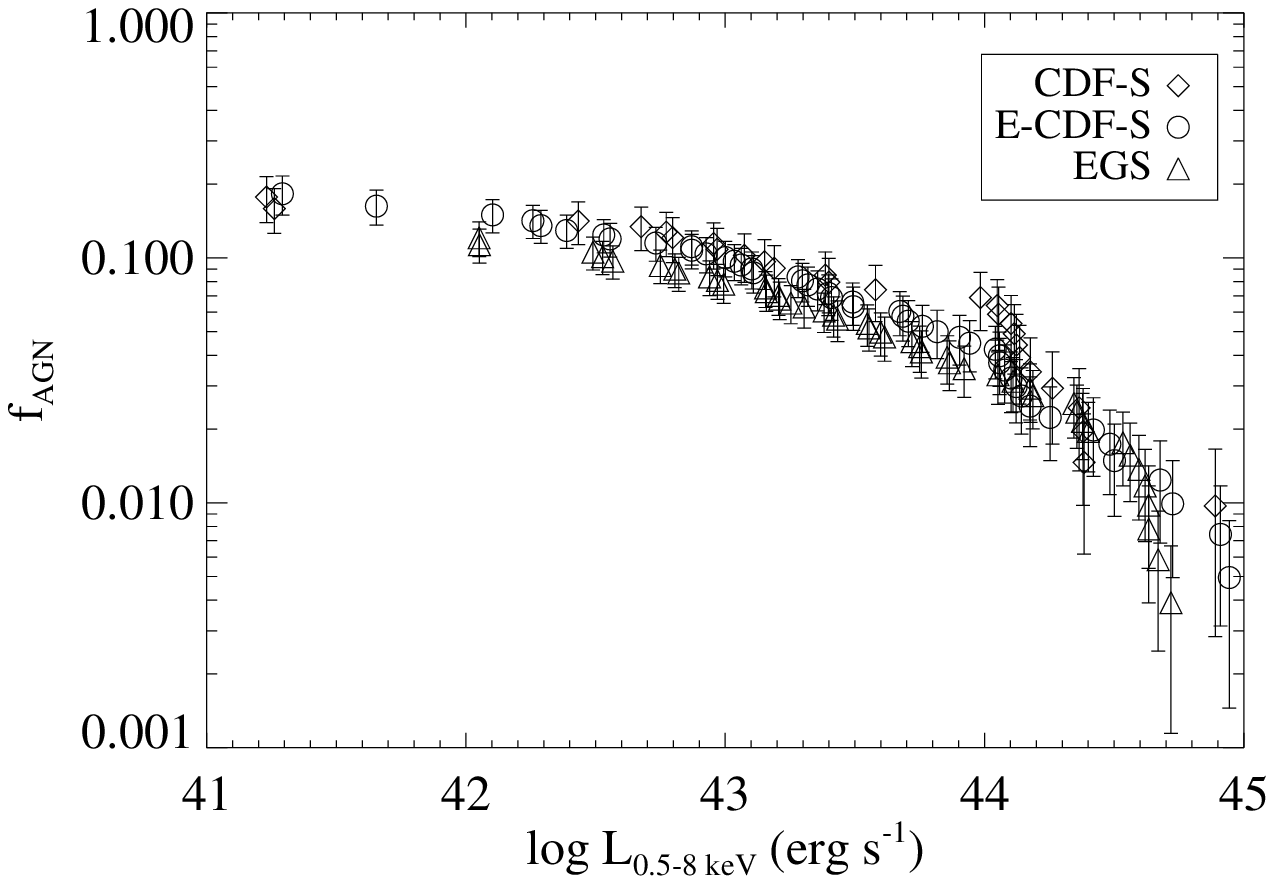}{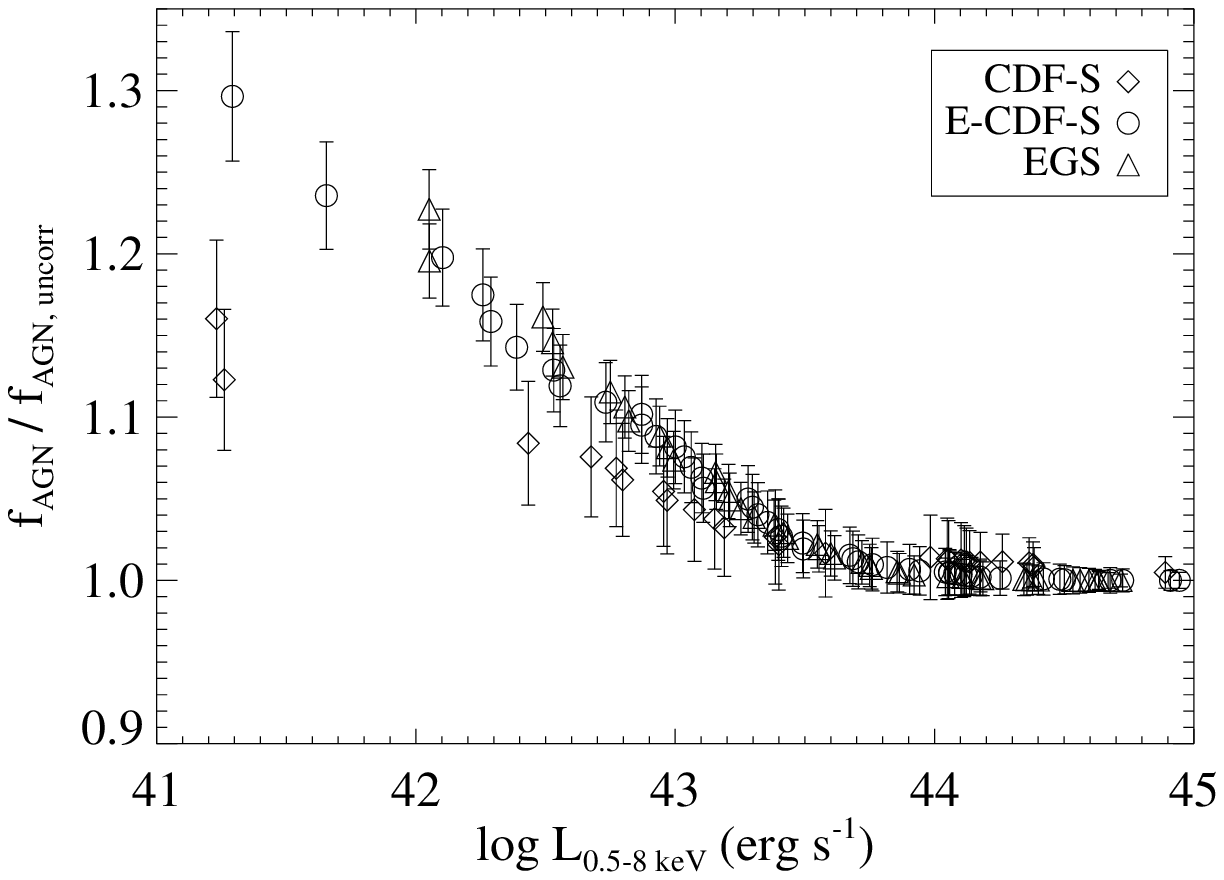}
\caption{\emph{Left:} The AGN fraction corrected for the \citet{tozz06} distribution of column densities ($f_{\rm AGN}$) as a function of the intrinsic, rest-frame 0.5--8~keV AGN luminosity. \emph{Right:} The ratio of the corrected-to-uncorrected AGN fraction as a function of the AGN luminosity. The fraction is plotted separately for each field, as indicated by the different symbols (note, however, that the \mbox{E-CDF-S} sources include those of the CDF-S). Errors are calculated following equation (\ref{E:agn_fraction_err}), but account only for the sampling error (i.e., $\sigma_{\rm phys}=0$ and $\sigma_{N_{\rm H}}=0$).  \label{F:fAGN_LAGN}}
\end{figure*}

To illustrate the effect of including the column density distribution in the calculation of the AGN fraction, we plot in Figure~\ref{F:fAGN_LAGN} the cumulative AGN fraction as a function of the AGN luminosity for the three fields. Our method of accounting for the column density distribution increases the AGN fraction overall, but particularly at lower redshifts and hence, on average, at lower AGN luminosities (by $\approx 10$--30\% at $L_{0.5-8.0\rm{~keV}} \sim 2 \times 10^{41}$~erg~s$^{-1}$). At the highest redshifts and AGN luminosities in our sample ($L_{0.5-8.0\rm{~keV}} \gtrsim 10^{44}$~erg~s$^{-1}$), the fraction is altered only slightly, since the emission probed by the \emph{Chandra} data is at higher rest-frame energies and hence less affected by the obscuration.  

Additionally, the correction produces larger changes in the shallower fields, such that the AGN fraction in the \mbox{CDF-S}, the field with the deepest X-ray data, has the least change and the EGS the greatest. This result is expected given that a larger fraction of high-column-density sources will be missed in the shallower X-ray fields at a given intrinsic AGN luminosity (and hence the AGN fraction will be biased low). We note, however, that, even after the correction is applied, the EGS tends to have the lowest cumulative AGN fraction at a given AGN luminosity (particularly at lower luminosities), suggesting that our adopted column-density distribution may not apply as well to the sources in the EGS as to those in the \mbox{E-CDF-S} (possibly due to cosmic variance) or that there may be some evolution in the cumulative AGN fraction with redshift (as the EGS data probe lower redshifts on average than the \mbox{E-CDF-S} data at a given AGN luminosity). Despite this issue, the cumulative fractions of the three fields are roughly consistent with one another given the uncertainties. Therefore, to obtain a larger sample size, we henceforth examine the AGN fraction of the combined \mbox{E-CDF-S} and EGS sample.

\section{Results}
As discussed in \S1, the AGN fraction, which gives the detection rate of AGNs in a given sample, is related to the duty cycle of AGN activity. Higher fractions imply that the AGNs in these systems spend more time in active states than do AGNs in systems with lower fractions. Therefore, the AGN fraction gives an indicator of the ubiquity of black-hole growth. Along with estimates of the relative levels of bulge and SMBH growth, this information can be used to understand how SMBHs grow during periods of vigorous starburst activity. In this section, we investigate the relation between SMBH growth and the SFR for our sample of starbursts, and we examine the dependence of the AGN fraction on the observed source properties. Properties of interest include the mid-infrared color (which gives a rough measure of the temperature of the emitting dust and thus the relative contributions from AGNs and star formation to the power source of the reprocessed emission), the rest-frame mid-infrared luminosity, and the star formation rate. 

For each of these properties, we use the combined \mbox{E-CDF-S} and EGS 70~\um\ sample and investigate two minimum rest-frame AGN cutoff luminosities: $L_{0.5-8.0\rm{~keV}} \ge 10^{41}$~erg~s$^{-1}$ and $L_{0.5-8.0\rm{~keV}} \ge 10^{43}$~erg~s$^{-1}$. However, due to the flux-limited nature of the \mbox{X-ray} and far-infrared surveys upon which our analysis is based, both this cutoff luminosity and the rest-frame mid-infrared luminosity and SFR are increasing functions of redshift (see Figure~\ref{F:z_dist}). Therefore, caution must be exercised when interpreting trends in the AGN fraction with luminosity or SFR when a large range in cutoff AGN luminosities is present (for reference, the cutoff AGN luminosities are indicated on the relevant plots). 

When examining the AGN fraction as a function of SFR, we used the subsample of 85 AGNs (and 999 70~\um\ sources in total) created by filtering out sources that fall below our adopted 70~\um\ S/N limit after subtracting the estimated contribution from the AGN to the 70~\um\ flux (see \S\ref{S:pred_AGN_fluxes}).  To avoid situations in which a single AGN dominates the fraction in a given bin, we construct the bins so that each contains a minimum of 10 AGNs and exclude any bin in which a single AGN contributes 30\% or more to the fraction in that bin (the contribution of each AGN to the total fraction depends on the distribution of limiting luminosities; see equation \ref{E:agn_fraction}). This method minimizes the effect on the fraction of a single AGN that might have, for example, an incorrect redshift estimate (such sources are expected to account for $\lesssim 10$\% of our sample; see \S\ref{S:redshifts}). Additionally, the AGN fraction is strictly valid only when calculated for complete samples (e.g., for all galaxies with $10<{\rm SFR}<30$~$M_{\odot}$~yr$^{-1}$ and redshifts less than the limiting redshift for a ${\rm SFR}=10$~$M_{\odot}$~yr$^{-1}$ galaxy). For our adopted minimum number of AGNs per bin (10), we found that the typical bin width is small enough such that the difference in limiting redshifts at the bin boundaries is much smaller than the typical limiting redshift of objects in that bin. Therefore, the samples in each bin should be roughly complete.

\subsection{The AGN Fraction and Mid-Infrared Color}
\begin{figure*}
\plottwo{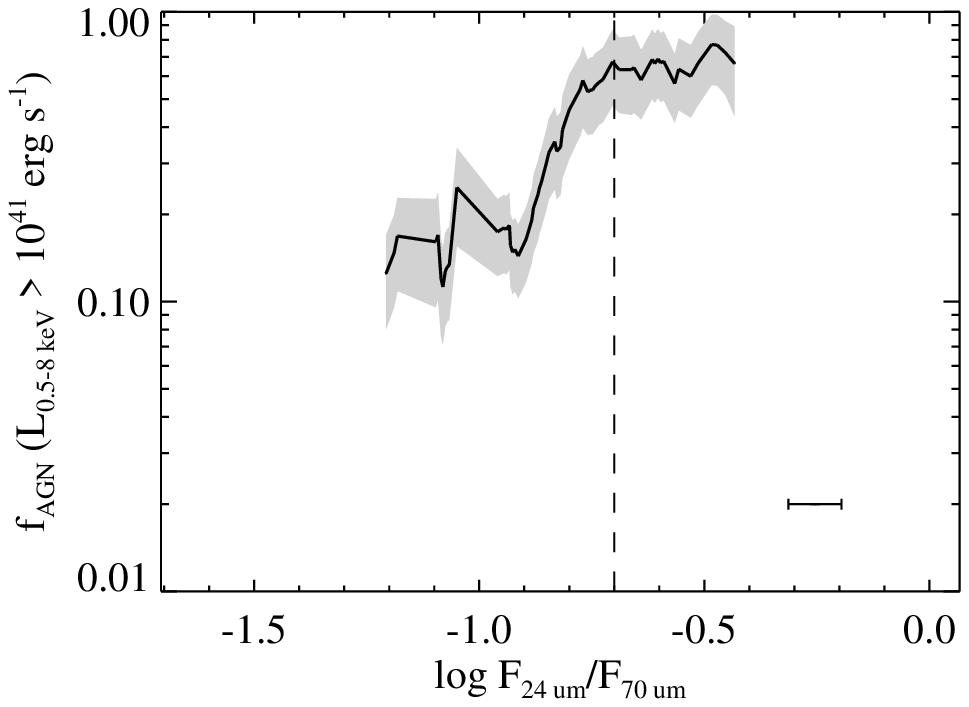}{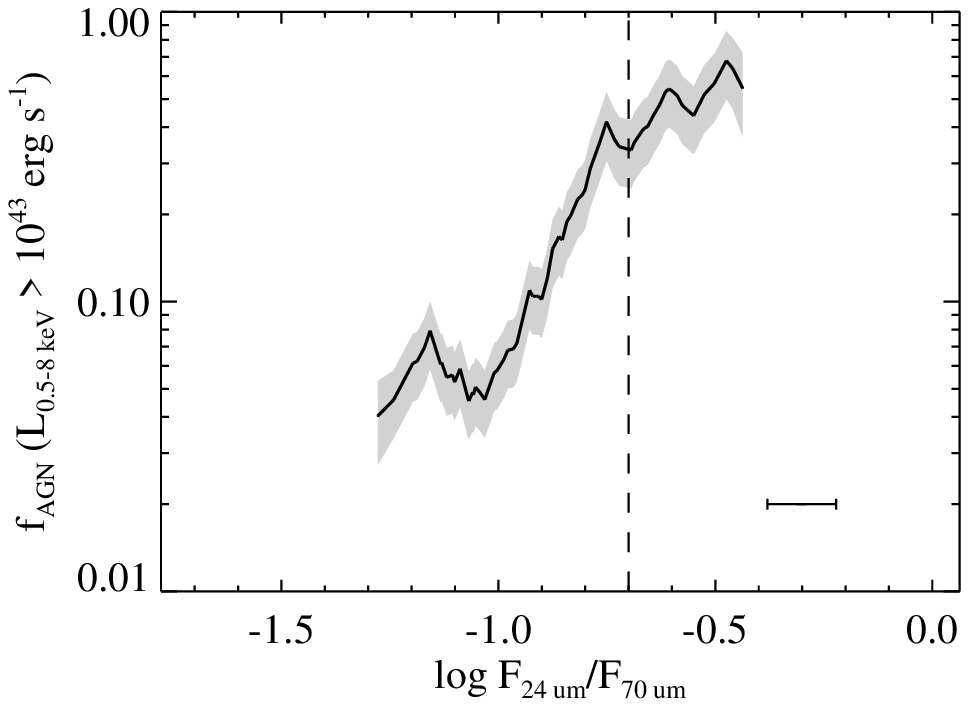}
\caption{The AGN fraction as a function of the mid-infrared color for AGNs with $L_{0.5-8.0\rm{~keV}} \ge 10^{41}$~erg~s$^{-1}$ (\emph{left}) and $L_{0.5-8.0\rm{~keV}} \ge 10^{43}$~erg~s$^{-1}$ (\emph{right}). The vertical, dashed line denotes $\log(F_{24}/F_{70}) \approx -0.7$. A sliding bin containing a minimum of 10 AGNs was used, the mean width of which is indicated in the lower right-hand corner. Variations in the fraction on scales smaller than this bin width are not significant. The shaded region indicates the 1-$\sigma$ errors.\label{F:fagn_mir_color}}
\end{figure*}

We begin by showing in Figure~\ref{F:fagn_mir_color} the AGN fraction as a function of the mid-infrared color of our 70~\um\ sources. It is clear from this figure that the fraction is a strong function of the mid-infrared color, rising from 5--10\% at the smallest values of the ratio $(F_{24}/F_{70})$ (indicative of cooler dust temperatures) to $\sim 60$--70\% at the largest values. The traditional dividing point of this ratio for $z\sim 0$ sources between dust powered by AGN-dominated emission and that powered by star formation-dominated emission is at $\log(F_{25}/F_{60}) \approx -0.7$  \citep[e.g.,][]{degr85,sand88}. This division also occurs at roughly the same ratio when \textit{Spitzer} bands are used (i.e., $\log [F_{24}/F_{70}] \approx -0.7$) and appears to hold out to at least $z\sim 1.5$ \citep[e.g.,][]{mull09}.  Indeed, at approximately this ratio, the AGN fraction appears to reach its maximum, implying that $\sim 60$--70\% of such sources host an AGN. Below $\log(F_{24}/F_{70}) \approx -0.7$, the fraction of sources of a given color that hosts an AGN falls rapidly as the color indicates cooler temperatures. Although care must be taken in interpreting the color at $z\gtrsim 1.5$ due to the increasing complexity of typical AGN and starburst spectra at the rest-frame wavelengths sampled by the 24~\um\ band in particular, this result extends the analyses of \citet{mull09} by showing that the mid-infrared color is a useful indicator of luminous AGN activity in the distant universe.

\subsection{The AGN Fraction and Mid-Infrared Luminosity}
\begin{figure*}
\plottwo{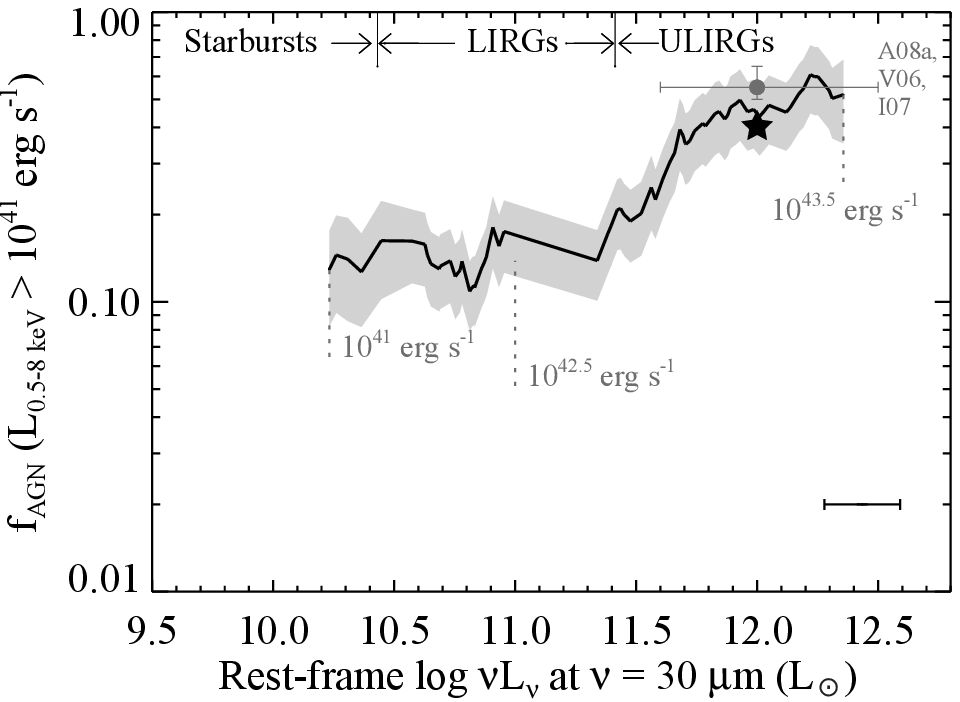}{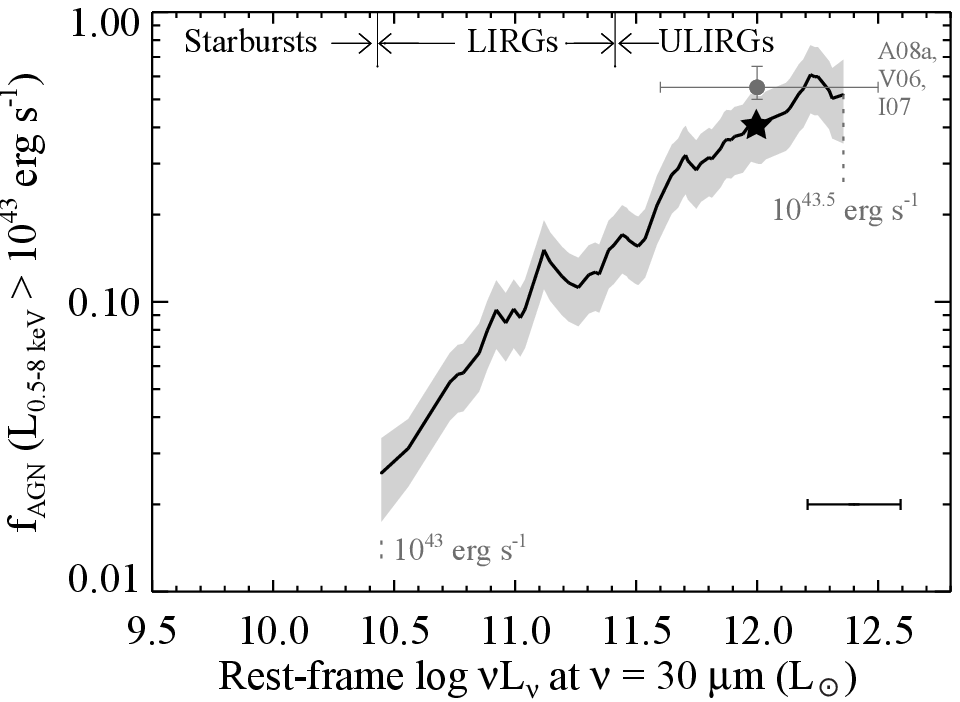}
\caption{The AGN fraction as a function of the rest-frame 30~\um\ luminosity for AGNs with $L_{0.5-8.0\rm{~keV}} \ge 10^{41}$~erg~s$^{-1}$ (\emph{left}) and $L_{0.5-8.0\rm{~keV}} \ge 10^{43}$~erg~s$^{-1}$ (\emph{right}). The divisions between starbursts, LIRGs, and ULIRGs were calculated using the starburst models of \citet{char01} by adopting a rest-frame luminosity range for LIRGs of $10^{11}$ $L_{\odot} < L_{\rm IR}<10^{12}$~$L_{\odot}$ and for ULIRGs of $L_{\rm IR}>10^{12}$~$L_{\odot}$. The approximate AGN fraction for local ULIRGs \citep{iman07,veil06,alex08_bh} is indicated by the circle and error bars, with the fraction of such systems found by \citet{alex08_bh} with $L_{2-10\rm{~keV}} \ge 10^{43}$~erg~s$^{-1}$ indicated by the star symbol.  The minimum AGN luminosity ($L_{0.5-8.0\rm{~keV}}$) in a number of bins is also indicated. The 1-$\sigma$ errors and mean bin size are indicated as in Figure~\ref{F:fagn_mir_color}. \label{F:fagn_L_30}}
\end{figure*}

In Figure~\ref{F:fagn_L_30}, we plot the AGN fraction against the rest-frame 30~\um\ luminosity of the source (note that this quantity includes any AGN contribution). It is clear from this figure that the fraction of sources hosting an AGN depends on the rest-frame mid-infrared luminosity, with higher fractions in sources with higher luminosities. The dependence becomes stronger when lower-luminosity AGNs are excluded, such that the fraction of sources hosting an AGN with $L_{0.5-8.0\rm{~keV}} \ge 10^{43}$~erg~s$^{-1}$ rises with the mid-infrared luminosity from a few percent at $L_{30} \approx 10^{10}$~$L_{\odot}$ to $\sim 60$\% at $L_{30} \approx 5 \times 10^{12}$~$L_{\odot}$. Therefore, more luminous mid-infrared sources, such as ULIRGs, are $\sim 10$ times more likely to host a luminous AGN than lower-luminosity sources, such as local starbursts. Such a result is to be expected if the AGN contributes to the mid-infrared emission, and the average contribution increases with increasing AGN luminosity (as is suggested by Figure~\ref{F:70um_comparison}).

The peak in the AGN fraction ($f_{\rm AGN} \sim 50$--60\%) in Figure~\ref{F:fagn_L_30} at the highest mid-infrared luminosities (corresponding broadly to ULIRG luminosities) is consistent with the fraction ($\approx 60$\%) of local ULIRGs at these SFRs identified by \citet{alex08_bh} as hosting \mbox{X-ray} AGNs or classified as Seyfert galaxies by \citet{veil06}. It is also consistent with the total AGN fraction of $\sim$ \mbox{50--65}\% implied by the detection rate of highly obscured AGNs (identified using \textit{Spitzer} spectroscopy) in a sample of local ULIRGs studied by \citet{iman07}. We note, however, that the luminosity cutoffs indicated in Figure~\ref{F:fagn_L_30} do not strictly correspond to LIRG and ULIRG cutoffs since, for example, cooler ULIRGs may have rest-frame 30~\um\ luminosities below the indicated cutoff. Furthermore, the differing selection criteria of these studies makes it likely that they probe AGNs of different luminosities on average (as noted earlier and indicated in Figure~\ref{F:fagn_L_30}, the minimum AGN luminosity to which we are sensitive varies with the redshift and hence rest-frame 30~\um\ luminosity). However, if we restrict the sample of \citet{alex08_bh} to those systems hosting an \mbox{X-ray} AGN with $L_{2-10\rm{~keV}} \ge 10^{43}$~erg~s$^{-1}$, which matches our selection criteria more closely, we find that 4 of 10 local ULIRGs host such an AGN, with an implied AGN fraction of $\sim 40$\%, in good agreement with our result. Therefore, our results suggest that the AGN fraction in the distant ULIRGs in our sample is similar to that of local ULIRGs and, furthermore, that the relation between AGN activity and star formation in these two populations is similar.

\subsection{The AGN Fraction and Star Formation Rate}\label{S:fagn_sfr}

\begin{figure*}
\plottwo{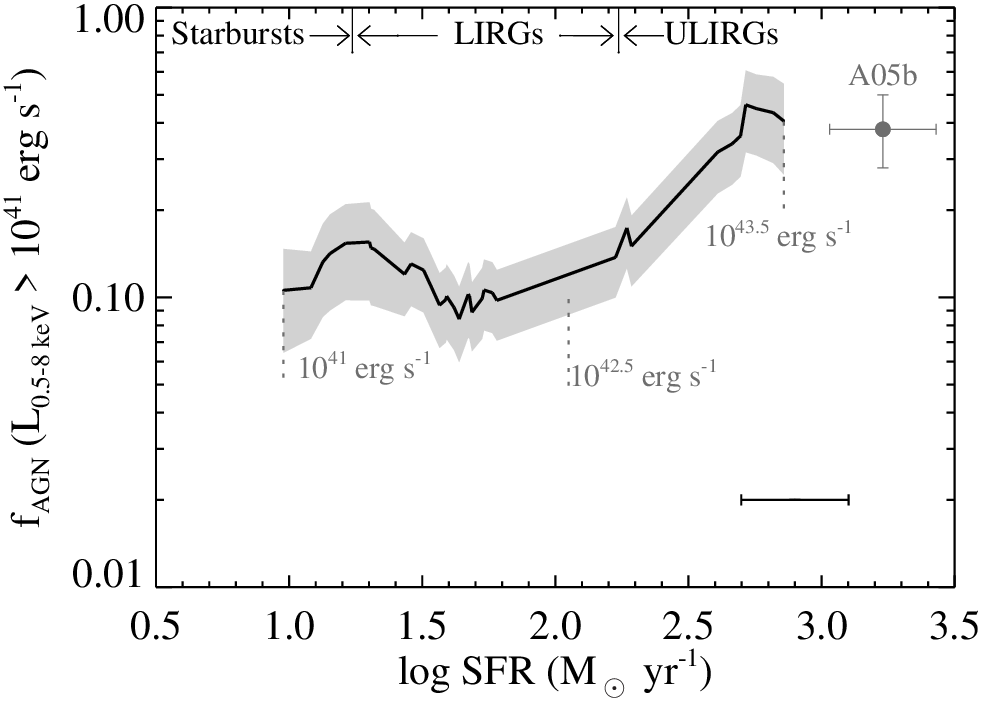}{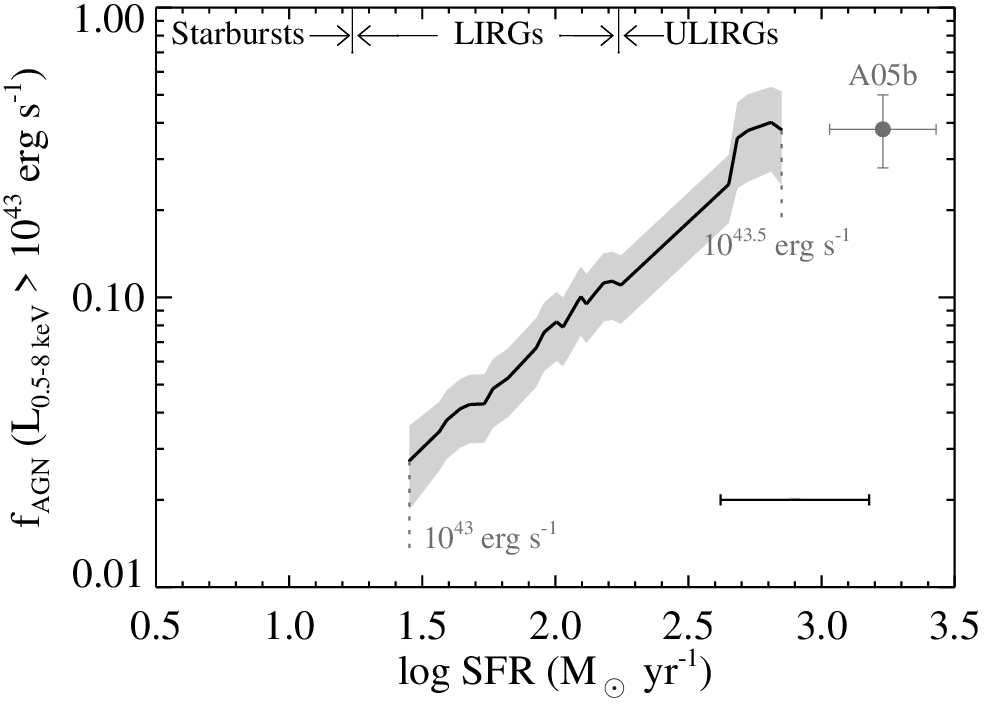}
\caption{The AGN fraction as a function of SFR for AGNs with $L_{0.5-8.0\rm{~keV}} \ge 10^{41}$~erg~s$^{-1}$ (\emph{left}) and $L_{0.5-8.0\rm{~keV}} \ge 10^{43}$~erg~s$^{-1}$ (\emph{right}). The approximate fraction for $z\approx 2$ submillimeter galaxies \citep{alex05b} is indicated. The minimum AGN luminosity ($L_{0.5-8.0\rm{~keV}}$) in a number of bins is also indicated. The 1-$\sigma$ errors and mean bin size are indicated as in Figure~\ref{F:fagn_mir_color}.\label{F:AGN_fraction}}
\end{figure*}

Figure~\ref{F:AGN_fraction} shows the cumulative AGN fraction for AGNs with $L_{0.5-8.0\rm{~keV}} \ge 10^{41}$~erg~s$^{-1}$  and $L_{0.5-8.0\rm{~keV}} \ge 10^{43}$~erg~s$^{-1}$ as a function of SFR.  Before calculating a SFR, we subtracted the expected AGN contribution from the observed 70~\um\ flux as described in \S\ref{S:pred_AGN_fluxes}. We then used the procedure described in \S\ref{S:SFRs} to infer a SFR from a given net 70~\um\ flux. It is clear from this figure that the AGN fraction in sources with lower SFRs (${\rm SFR}\sim 10$~$M_{\odot}$~yr$^{-1}$) depends strongly on the AGN luminosity ($L_{0.5-8.0\rm{~keV}}$). When lower-luminosity AGNs are included, the fraction of sources hosting an AGN at low SFRs is quite high at \mbox{$\sim 10$--20\%}.  When only higher-luminosity AGNs are considered ($L_{0.5-8.0\rm{~keV}} \ge 10^{43}$~erg~s$^{-1}$), the AGN fraction at low SFRs is much lower, rising from $\approx 4$--5\% at SFRs of roughly 10~$M_{\odot}$~yr$^{-1}$ to $\sim 40$\% at the highest SFRs as $f_{\rm AGN} \propto {\rm SFR}^{0.75}$. 

Additionally, SMGs, which generally lie at high redshift ($z\sim 2$) where the reprocessed emission from cool dust peaks at observed-frame submillimeter wavelengths, have larger average SFRs than most of the sources present in our sample. Their typical SFRs are on the order of 1000--3000~M$_{\odot}$~yr$^{-1}$, and the reprocessed emission in these systems appears to be largely powered by star formation \citep[e.g.,][]{alex05,mene07,vali07,pope08}. SMGs also appear to have a high AGN fraction \citep[$\sim 20$--50\%,][]{alex05b, lair10}, and are thought to be the short-lived phase of rapid accretion during a massive merger. Their relation to local ULIRGs is unclear \citep[e.g.,][]{saji07,syme09}, but the two classes appear to share many of the same properties, with SMGs being roughly scaled up by an order of magnitude in far-infrared luminosity and SFR \citep{saji07,vali07}. We note that although our sample does not contain a sufficient number of ${\rm SFR} \gtrsim 1000$~M$_{\odot}$~yr$^{-1}$ objects to determine a reliable AGN fraction in this regime, extrapolation of our determination of the AGN fraction (Figure~\ref{F:AGN_fraction}) to SFRs of $\sim 2000$~M$_{\odot}$~yr$^{-1}$ results in an AGN fraction roughly consistent with the fractions found by \citet{alex05b} and \citet{lair10} for SMGs using independent methods and a different field (the CDF-N).

\begin{figure*}
\plottwo{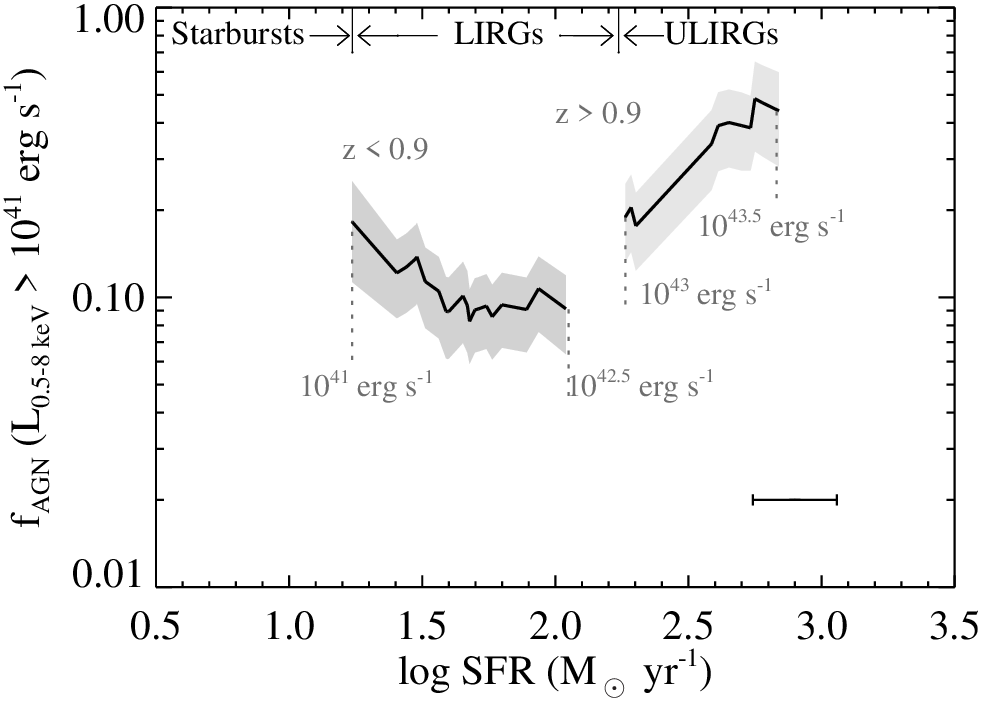}{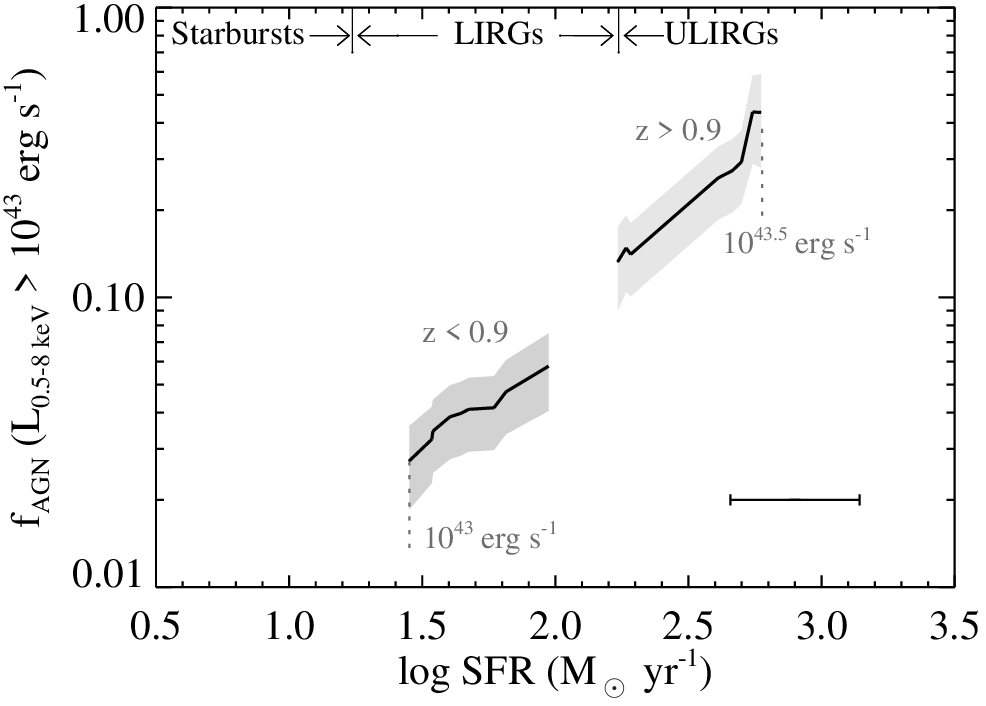}
\caption{The same as Figure~\ref{F:AGN_fraction} but for two subsamples: sources with $z<0.9$ (\textit{dark gray}) and those with $z>0.9$ (\textit{light gray}). \label{F:AGN_fraction_zbin}}
\end{figure*}

To investigate whether the AGN fraction has a dependence on redshift, we plot in Figure \ref{F:AGN_fraction_zbin} the AGN fraction against the SFR for two subsamples divided by the median redshift of the AGN sample, $z=0.9$. Due to the sensitivity limits of the FIDEL survey, at high redshifts only the most luminous 70~\um\ sources (with correspondingly high SFRs) are detected. Conversely, at low redshifts, no very luminous 70~\um\ sources are present in the survey fields. Due to this effect, the two subsamples do not have sufficient overlap to judge whether there is any dependence of the AGN fraction with redshift at a given SFR. Additionally, as discussed earlier, the minimum AGN luminosity for which the cumulative fraction is calculated changes with the SFR, as indicated in Figure \ref{F:AGN_fraction_zbin}. This changing minimum is largely responsible for the trend of decreasing AGN fraction with increasing SFR seen at $z<0.9$ in the left panel of Figure \ref{F:AGN_fraction_zbin}. Two opposing effects sum to create this trend: (1) the number of lower-luminosity AGNs per bin decreases with increasing SFR due to a steady rise in the minimum cutoff AGN luminosity with increasing redshift (and hence SFR) and (2) the contribution to the cumulative fraction from higher-luminosity AGNs increases with the SFR (a trend visible in the right panel). Overall, the net effect is a gradual decrease in the AGN fraction with increasing SFR at $z<0.9$. At $z>0.9$, the change in the cutoff AGN luminosity across the sampled range of SFR is small, and should not have a large effect on the overall trend.

To examine the effects that AGNs with warm mid-infrared colors (i.e., those sources that may have large AGN contributions to their 70~\um\ fluxes) have on our results, we have further filtered our sample to exclude all sources with $\log(F_{24}/F_{70}) \gtrsim -0.7$. The only significant effect of this filtering is reduce the AGN fraction overall (and to reduce the sample such that a smaller range of SFRs is probed). The same trends are present both with and without this filtering.

\subsection{Black-Hole and Bulge Growth}
\begin{figure*}
\plottwo{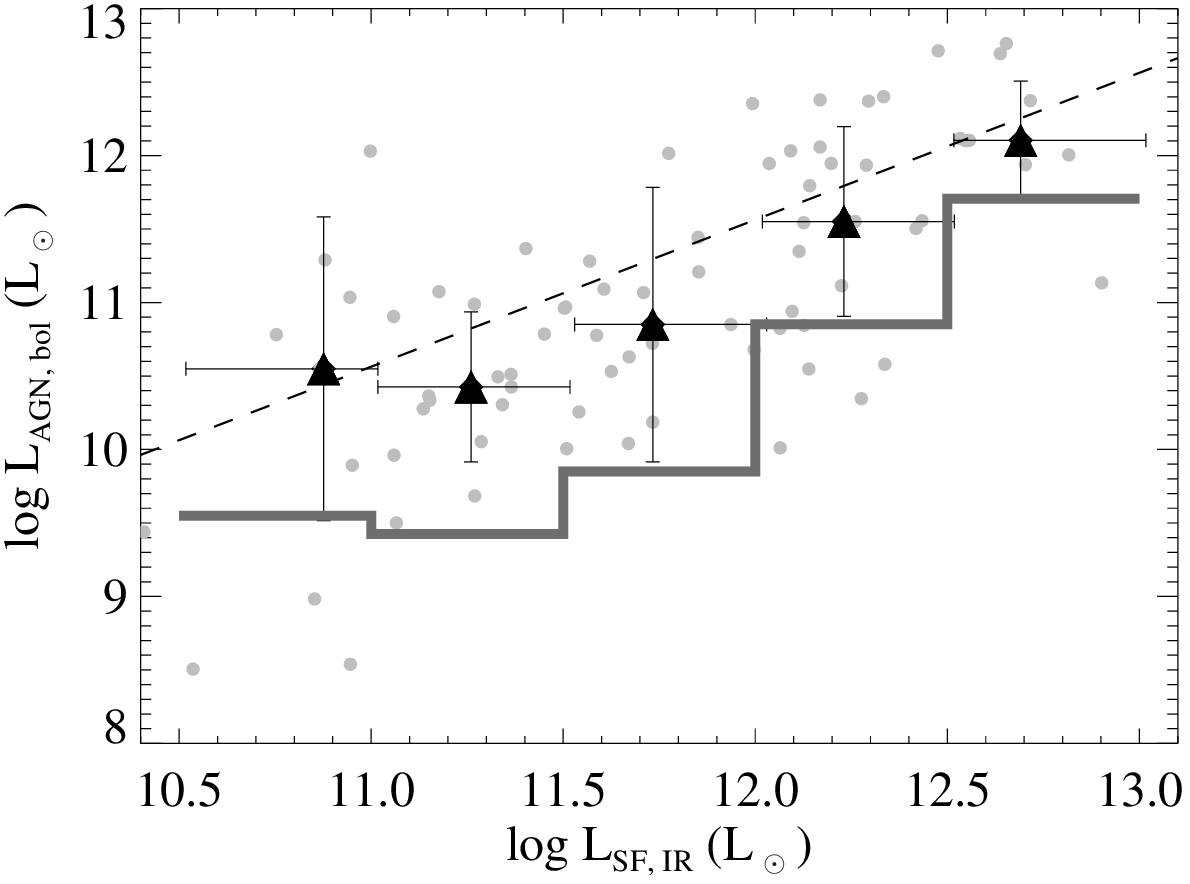}{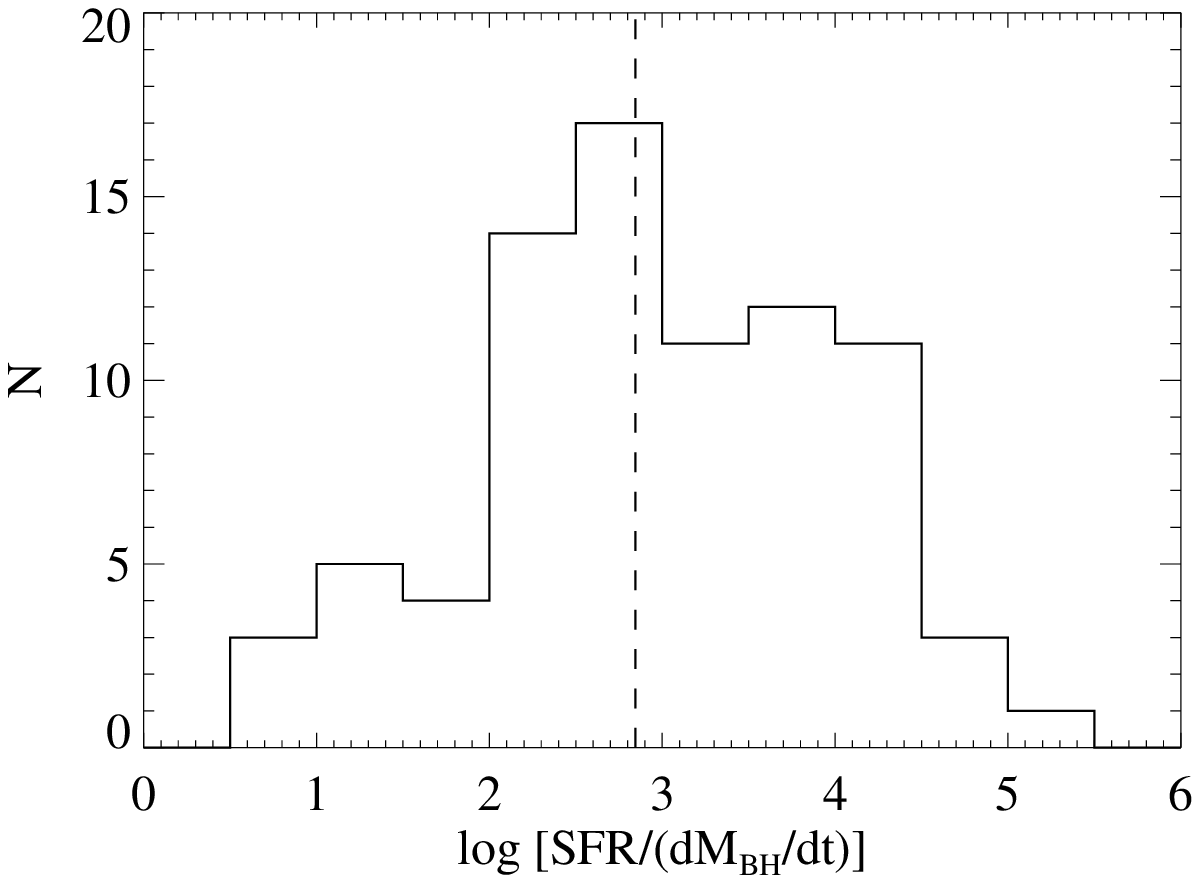}
\caption{\textit{Left:} The bolometric AGN luminosity, $L_{\rm AGN,~bol}$, versus the 8--1000~\um\ luminosity due to star formation, $L_{\rm SF,~IR}$. Gray points show individual systems, filled diamonds indicate the median of $L_{\rm AGN,~bol}$ of samples binned on $L_{\rm SF,~IR}$. For the binned points, horizontal error bars indicate the bin sizes, and vertical error bars indicate the standard deviation. The solid line indicates the lower limit on the time-averaged $L_{\rm AGN,~bol}$, after adjusting $L_{\rm AGN,~bol}$ by the average cumulative AGN fraction in each bin. The dashed line indicates the relation expected from the local ratio of bulge mass to black-hole mass found by \citet{hari04}. \textit{Right:} Histogram of the ratio of SFR to black-hole accretion rate. As in the left panel, the dashed line indicates the average ratio expected from the \citet{hari04} relation if the SFR is equal to the bulge growth rate. \label{F:Lagn_vs_Lsf}}
\end{figure*}

It is now well established that black holes and their host bulges are intimately connected \citep[e.g.,][]{ferr_ford05}. To illustrate how the growth rates of the AGNs in our sample compare to those of their bulges, we binned our AGN sample by SFR and calculated the mean bolometric AGN luminosity (calculated following \S\ref{S:corrections}) in each bin. In this comparison, we implicitly assume that the bulge growth rate is approximated by the total SFR, although this may not be strictly true (e.g., in late-type galaxies). In the left panel of Figure~\ref{F:Lagn_vs_Lsf}, we compare these luminosities to the \mbox{8--1000}~\um\ luminosities from star formation derived following equation~(\ref{E:kennsfr}). 

It appears that, in systems identified as having AGNs, the SMBHs and bulges in our sample are growing concurrently on average, across a wide range of SFR (roughly two orders of magnitude), and at relative rates that would produce or maintain the scaling observed locally \citep[calculated using the conversions described above and assuming ${\rm SFR}/\dot{M}_{\rm BH} \propto M_{\rm bulge}/M_{\rm BH} \approx 700$; e.g.,][]{hari04}. We note that we are likely missing many systems with low SFRs that would fall mostly in the lowest bin. Depending on the average AGN luminosity of such systems, our value for the average AGN luminosity in this bin could be biased either high (if the missed systems have weak AGN activity on average) or low (if they have strong AGN activity on average). The effect of systems in the latter category (with high $L_{\rm AGN,~bol}$ and low $L_{\rm SF,~IR}$) will be reduced somewhat due to the global increase in specific SFR with redshift, which will tend to reduce the fraction of luminous AGNs that lack significant star formation. Nevertheless, the uncertainty in the lowest bin is large and its value should be treated with caution.

However, because many galaxies at a given SFR lack luminous AGN activity, the average AGN luminosity (and hence SMBH growth rate) across all galaxies in a given bin will be somewhat lower. We can estimate this average AGN luminosity by multiplying through by the appropriate AGN duty cycle (traced by the AGN fraction shown in Figure~\ref{F:AGN_fraction}). This estimate is shown by the solid line in the left panel of Figure~\ref{F:Lagn_vs_Lsf} and is well below that expected from the local scaling relation (with the exception of the lowest-SFR bin, which, as noted above, is likely strongly affected by incompleteness), implying that the bulges in our sample could be growing at a faster rate on average relative to their SMBHs than expected from the local relation. However, we note that the bulge growth rate will often be significantly less than the total SFR of the galaxy and, for this reason (among others discussed in the next section), this average AGN luminosity should be interpreted as an approximate lower limit to the true one. 

Additionally, in the right panel of Figure~\ref{F:Lagn_vs_Lsf}, we show the distribution of the ratio of the SFRs of the host galaxies to the SMBH accretion rates calculated in \S\ref{S:corrections}. Again, it is clear that the ratio of SFR to accretion rate for our sample of AGNs is broadly consistent with that expected from the median observed ratio of bulge mass to black-hole mass found locally, although our sources tend to lie at slightly higher ratios (median ${\rm SFR}/\dot{M}_{\rm BH} \approx 1000$). However, given the uncertainties in the calculation of these quantities, our results agree well with expectations from the $M_{\rm BH}$-$M_{\rm bulge}$ relation and studies of the volume-weighted bulge and SMBH growth in local galaxies \citep[e.g.,][]{heck04}.  

In both panels of Figure~\ref{F:Lagn_vs_Lsf}, there is large scatter about the local scaling. Much of this scatter may be due to the systematics discussed above, but could also indicate that some of the systems are undergoing large deviations (by factors of up to $\sim 100$) from concurrent SMBH-bulge growth. 

\section{Discussion}\label{S:AGN_activity}
In the previous section, we presented the results of an analysis of SMBH growth in distant luminous starbursts. We investigated the AGN fraction as a function of a number of physical properties and examined the relative growth rates of the SMBHs and their host galaxies. In particular, we found a strong dependence of the AGN fraction on the SFR in our sample, with the fraction rising to $\sim 30$\% at the highest SFRs, implying a high duty cycle of luminous AGN activity in such systems.

However, as discussed previously, due to the sensitivity limits of the \mbox{X-ray} data and the fact that high-SFR objects tend to lie at higher redshifts, we are not generally sensitive to lower-luminosity AGNs in these systems. Therefore, the total cumulative fraction of AGNs in intermediate and high-SFR systems may be higher than shown in Figure~\ref{F:AGN_fraction}. Additionally, as discussed in \S\ref{S:obscured_AGNs}, it is likely that we have missed a significant population of AGNs with very high column densities due to their weak \mbox{X-ray} emission. Although we have attempted to correct for this missing population using an estimate of the distribution of intrinsic AGN column densities, our adopted distribution \citep[based on that of\ ][]{tozz06} may not apply well to high-SFR objects. Such objects, which presumably harbor large amounts of cold gas, may have higher-than-average values of the column density \citep[e.g.,][]{alex05}. Therefore, we may underestimate the AGN fraction in such sources. 

To assess how sensitive our results are to changes in the population of Compton-thick sources, we altered the $N_{\rm H}$ distribution described in \S~\ref{S:calc_AGN_fraction} by maintaining a flat distribution out to $N_{\rm H}=10^{25}$~cm$^{-2}$, effectively increasing the number of Compton-thick sources by a factor of $\approx 3.5$ \citep[such an increase is within the range of some recent predictions; e.g.,][]{fior09}. With this change, we find that the AGN fraction increases at all SFRs by $\sim 30$\%, with a somewhat larger increase at low SFRs or luminosities due to the larger effect of obscuration on the observed-frame \mbox{X-ray} fluxes at low redshifts (which generally correspond to lower SFRs and luminosities). Therefore, depending on the true number of Compton-thick sources, it is conceivable that our determination of the AGN fraction could be low by $\sim 30$\% or more.

A related issue is that of incompleteness due to statistical effects that results in underestimates of the number of sources near the flux limit. As noted in Section~\ref{S:sample}, simulations indicate that the incompleteness of our 70~\um\ sample is small at $S/N=3$, our adopted cutoff. As a test of this assertion, we performed the entire analysis again, but with a cutoff $S/N=5$.  With this cutoff, our sample becomes almost a factor of two smaller (556 70um sources in total versus 1022 sources for $S/N=3$), but the trends we observe in the AGN fraction do not change appreciably, except for the trend with the SFR for higher-luminosity AGNs ($L_{0.5-8 keV} > 10^{43}$~erg~s$^{-1}$); see Figure~\ref{F:AGN_fraction}, right). In this case, the overall trend is very similar, but the overall normalization is lower by 20--30\% (although they are still consistent with the 1-$\sigma$ errors). Since these higher-luminosity AGNs tend to lie at higher redshift, the removal of the fainter sources has a larger effect on this sample than on the sample of lower-luminosity AGNs (those with $L_{0.5-8 keV} > 10^{41}$~erg~s$^{-1}$), as higher-redshift sources are fainter on average. Therefore, evolutionary effects could explain the lower normalization if the AGN fraction at a given SFR has a dependence on redshift. Alternatively, systematic errors in the estimated AGN contribution that lead to misestimates of the AGN contribution at shorter rest wavelengths or higher AGN luminosities (both of which are relevant for the high-redshift sources removed) could contribute to the observed effect. Unfortunately, our data do not allow us to  distinguish between these two possibilities, but we note that longer-wavelength far-infrared observations, which would suffer less from AGN contamination even at high redshifts, would be useful in this regard.

When we examine the growth rates of the SMBHs and galaxies in our sample, we find that the ratio of SMBH growth rate to bulge growth rate for the systems with identified AGNs agrees well on average with that expected from the local scaling relation. However, when we include our estimates of the AGN duty cycle, the average SMBH growth rate is a factor of 3--10 lower (neglecting systems with low SFRs where incompleteness likely biases our result significantly), suggesting that the bulges could be growing faster relative to the SMBHs than expected from the local scaling. While this average SMBH growth rate should be considered a lower limit for the reasons outlined above, our results are consistent with \citet{merl06}, who derived estimates of the evolution of the total star formation and accretion density of the Universe. Merloni et al.\ found that the accretion rate of SMBHs falls more quickly than the SFR from $z\sim 1$ to the present day. This finding implies that the SMBHs were more massive relative to their bulges at this and higher redshifts than they are in the local universe. Therefore, to reproduce the local $M_{\rm bulge}/M_{\rm BH}$ relation, the bulges must grow faster relative to their SMBHs in the distant universe. Our estimates of the ratio of SMBH to bulge growth rate, corrected for the AGN duty cycle at a given SFR, are consistent with this scenario.

Lastly, although a direct connection between AGN activity and star formation is appealing, the SFR may not be the fundamental quantity that drives the AGN fraction. In particular, the stellar mass of a galaxy is known to be related to the likelihood of AGN activity, with more massive galaxies on average being more likely to host an AGN. Additionally, simple evolution of the AGN fraction with redshift in all galaxies could account for our results, since our sample is largely degenerate between redshift and SFR. Recently, \citet{brus09} studied the dependence of the AGN fraction on the stellar mass for a sample of \mbox{X-ray} selected AGNs in the \mbox{CDF-S} and found that, over a redshift range similar to that of our study, the AGN fraction increases from a few percent at masses of $\sim 10^{10}$~M$_{\odot}$ to $\sim 30$\% at masses of $\gtrsim 3\times 10^{11}$~M$_{\odot}$. This trend is roughly consistent with the trend we see with SFR and, consequently, either property could be the fundamental driver of the AGN fraction. Additionally, \citet{xue10} found evidence that the AGN fraction shows a strong dependence on the stellar mass, as well as some dependence on redshift at a given SFR. Unfortunately, due to the difficulties in the derivation of stellar masses for luminous AGN hosts (due to the presence of significant, sometimes even dominant, emission from the AGN at the optical and near-infrared wavelengths typically used for stellar-mass estimation) and our relatively small and flux-limited sample (which results in a degeneracy of redshift and SFR), a detailed investigation of these questions is beyond the scope of this paper.  

In summary, given the various uncertainties discussed above, our findings on the total AGN fraction in distant luminous starbursts are broadly consistent with studies that have examined the ULIRG regime locally and the SMG regime at $z\sim2$. Our results indicate, when the SFR is assumed to be the primary driver of AGN activity, that the fraction of systems hosting a higher-luminosity AGN increases strongly with the SFR and that SMBHs and bulges grow together on average, implying an intimate connection between SMBH accretion and star formation during periods of vigorous growth. 

\section{Conclusions and Future Work}
We present an analysis of AGN activity in a sample of starburst galaxies in the \mbox{E-CDF-S} and EGS fields. The sample was constructed from the FIDEL mid-infrared \textit{Spitzer} survey that traces the reprocessed emission from young stars and AGNs. Our sample is roughly complete for SFRs above  $\sim 100$~$M_{\odot}$~yr$^{-1}$ (${\rm SFR} \sim 600$~$M_{\odot}$~yr$^{-1}$) to $z=1.0$ ($z=2.0$). Using this sample, we investigate how the incidence and strength of AGN activity, identified using the deep \textit{Chandra} data for these fields, relates to the physical properties of the galaxies. In particular, we find that the fraction of sources that host an AGN depends strongly on the source's mid-infrared color, rest-frame mid-infrared luminosity, and, especially at higher intrinsic AGN luminosities, on the SFR of the galaxy. 

The dependence of the AGN fraction on the mid-infrared color confirms that the AGNs contribute significantly to the heating of the dust that emits in the mid-infrared. At warmer colors, approximately above the fiducial value of $\log (F_{\rm 24 \mu m}/F_{\rm 70 \mu m}) >-0.7$  \citep[e.g.,][]{mull09}, at least $\sim 70$--80\% of such sources host an AGN in the distant universe. The fraction of sources with an AGN decreases steadily with decreasing dust temperature, an indication that systems with cooler dust generally lack a luminous AGN and therefore their mid-infrared emission is due primarily to reprocessed emission from young stars.

A related effect is the rise in the AGN fraction with the rest-frame mid-infrared luminosity, which likely suggests again that the AGN contributes a greater fraction of the emission that heats the dust at higher mid-infrared luminosities. At the highest rest-frame mid-infrared luminosities sampled by our study (corresponding approximately to ULIRG luminosities, although the selection is somewhat different), the AGN fraction rises to $\sim 60$--80\%, in rough agreement with the AGN fraction determined for local ULIRGs by a number of recent studies using a variety of methods \citep[e.g.,][]{veil06,iman07,alex08_bh}.

Star formation also often plays an important role in powering the mid-infrared emission. Therefore, we have estimated the contribution from the AGN to the mid-infrared luminosity using empirical AGN SEDs, thus allowing us to disentangle the relative contributions from the AGN and star formation to this luminosity. After accounting for the AGN-powered emission and filtering our sample to avoid biases, we find that star formation appears to power the bulk of the mid-infrared emission in the remaining sample and that luminous AGN activity is more common in systems with higher SFRs (with the caveat that, due to the nature of our sample, the SFR, mass, and redshift of our sources are largely degenerate). At the highest SFRs ($\sim 1000$~$M_{\odot}$~yr$^{-1}$), the fraction of sources with an AGN rises to $\sim 30$--40\%. This fraction is roughly consistent with that derived for high-redshift SMGs \citep{alex05b,lair10}. At lower SFRs ($\sim 30$~$M_{\odot}$~yr$^{-1}$), the fraction of sources with a luminous AGN ($L_{0.5-8.0\rm{~keV}} \ge 10^{43}$~erg~s$^{-1}$) falls to a few percent. However, when lower-luminosity AGNs ($L_{0.5-8.0\rm{~keV}} \ge 10^{41}$~erg~s$^{-1}$) are included, the fraction is $\approx 10$\% for ${\rm SFR} \lesssim 100$~$M_{\odot}$~yr$^{-1}$. This detailed relation between the AGN fraction, and hence AGN duty cycle, and the SFR should provide useful constraints on large-scale models of galaxy and SMBH evolution \citep[e.g.,][]{dima08,hopk08}. Such models should reproduce AGN duty cycles that are consistent with the results presented here.

Lastly, we made rough estimates of the growth rates of the SMBHs and bulges in our sample. We found that, for systems with detected AGN activity, the median ratio of bulge to SMBH growth is consistent with that expected from the local scaling relation, although with large scatter. This result implies that the SMBHs and bulges in these systems are growing concurrently on average, even during periods of intense star formation, at relative rates that would produce or maintain the scaling observed locally. However, we do find a large scatter in this ratio, suggesting that in individual systems there are periods of rapid SMBH growth that are unaccompanied by rapid bulge growth (and vice-versa), although systematic uncertainties may account for much of this scatter. When the AGN duty cycle is included, the lower limit on the average ratio of SMBH-to-bulge growth across all systems (not only those with detected AGN activity) suggests that the bulges in these distant luminous starbursts could be growing more quickly relative to their SMBHs than expected from the local scaling relation, consistent with recent predictions of the evolution of the total star formation and accretion density of the Universe \citep{merl06}.

In summary, our results demonstrate a close connection between AGN activity and star formation in distant starbursts and suggest that SMBHs and their bulges grow together on average over a wide range of growth rate. However, a great deal of further work is required to address a number of remaining issues. For instance, much uncertainty remains in the determination of the AGN contribution to the mid-infrared flux, a critical step in estimating the SFRs. Deep, longer-wavelength data (up to $\sim 500$\um) from the \textit{Herschel} telescope, which will primarily trace cool dust emission not associated with the AGN, should prove invaluable in both increasing sample sizes and accurately estimating SFRs of sources at high redshift. Near-future hard \mbox{X-ray} observatories, such as NuSTAR, will be helpful for the identification of highly obscured AGNs. Lastly, improved understanding of the relation between the observed \mbox{X-ray} emission and the bolometric luminosity of AGNs \citep[e.g.,][]{vasu07} will also be very helpful in constraining the AGN mid-infrared contribution. 

\acknowledgments{Support for this work was provided by NASA through Chandra Awards SP8-9003A (DAR, WNB, YX, BL)  and SPO8-9003B (FEB) issued by the Chandra \mbox{X-ray} Observatory Center, which is operated by the Smithsonian Astrophysical Observatory. We also acknowledge NASA ADP grant NNX10AC99G (DAR, WNB, YX), the Royal Society (DMA), and a Philip Leverhulme Prize (DMA) for support. We also thank M.\ Dickinson for helpful feedback and A.\ Goulding and J.\ Mullaney for help in interpreting the AGN SEDs and mid-infrared color ratios. 

\bibliographystyle{apj}
\bibliography{/home/rafferty/Documents/Bibliography/master_references}

\appendix
\section{Photometric Redshifts for sources in the \mbox{E-CDF-S} and CDF-N}
A number of photometric redshift catalogs are available for the \mbox{E-CDF-S} \citep[e.g.,][]{wolf04,graz06} that have proved very useful to studies such as ours of the average properties of large samples of galaxies. Recently, however, new optical, UV, and near-to-mid-infrared data have become available for the \mbox{E-CDF-S}, as well as the \mbox{CDF-N}. These data should allow derivation of photometric redshifts that are generally of improved quality and that additionally include sources with fainter fluxes. To supplement the spectroscopic redshifts used in the work described in this paper and others \citep[e.g.,][]{xue10}, we have produced photometric redshift catalogs for nearly all detected optical sources in the 2~Ms \emph{Chandra} Deep Fields (the \mbox{CDF-S} and \mbox{CDF-N}) and the 250~ks \mbox{E-CDF-S}. Although we do not use photometric redshifts for the \mbox{CDF-N} in this paper, we include them here for completeness and ease of reference. In this appendix, we briefly describe our method of deriving photometric redshifts and present estimates of their quality by comparing to spectroscopic redshifts in these fields (further details are given with the catalog).\footnote{The same method was also used to derive the photometric redshifts for the EGS used in this paper, but we have not created a catalog of all optical sources. The reader is referred to \S\ref{S:redshifts} for a discussion of the quality of the photometric redshifts (including those from the EGS) for the 70~\um\ sample presented in this paper. We note, however, that a comparison of our EGS photometric redshifts with the photometric redshifts derived by \citet{alme09} for 96 \mbox{X-ray} and mid-infrared identified AGNs shows that our photometric redshift estimates are of comparable quality. }

\subsection{Photometric Catalogs}
For the \mbox{E-CDF-S}, we constructed photometric catalogs using the following catalogs: the MUSYC BVR-detected optical catalog \citep{gawi06}, the COMBO-17 optical catalog \citep{wolf04,wolf08}, the GOODS-S MUSIC catalog \citep{graz06}, the MUSYC near-infrared catalog \citep{tayl09}, the SIMPLE \emph{Spitzer} IRAC catalog \citep{dame09}, the GALEX UV catalog (NUV and FUV) from the GALEX Data Release 4, and the GOODS-S deep $U$-band catalog \citep{noni09}. In the CDF-N, the following catalogs were used: the GOODS-N HST ACS and \emph{Spitzer} IRAC photometric catalogs \citep{dick03}, the \mbox{CDF-N} \emph{Spitzer} IRAC catalog derived from unpublished IRAC archival data, the GALEX HDF-N deep imaging survey catalog from the GALEX Data Release 4, and the ACS GOODS-N region $K_s$ ($<24.5$) catalog \citep{barg08}. The sources were cross matched by position using a matching radius of 0\farcs5--1\arcsec\ (depending on the positional uncertainty of the catalogs). The final \mbox{E-CDF-S} photometric catalog comprises a total of 105,825 unique sources, and the \mbox{CDF-N} catalog comprises 48,858 sources.

\subsection{Galaxy, Hybrid, and Stellar Templates}
To model the galaxies, the 259 PEGASE galaxy templates used by \citet{graz06}\footnote{Provided with the EAZY distribution. See \url{http://www.astro.yale.edu/eazy/}.} that span a wide range of star formation history and intrinsic extinction were used. To model quasars and galaxy-AGN hybrids, the 10 AGN templates of \citet{poll07}\footnote{See \url{http://cass.ucsd.edu/SWIRE/mcp/templates/swire_templates.html}.} that include a variety of empirical quasar and Seyfert templates were used. Additionally, we constructed a set of AGN-galaxy hybrid templates by combining 5 typical AGN templates (type-1, type-2, QSO 1, and QSO 2) and 16 typical galaxy templates (elliptical, spiral, and starburst) with a variety of relative contributions. For each pair of AGN and galaxy templates, the templates were normalized by the total integrated flux and four hybrid templates with varying AGN/Galaxy ratios (90:10, 75:25, 50:50, and 25:75) were produced. A total of 330 AGN/galaxy-hybrid templates were used ($330=5\times 16\times 4+10$). 

In addition to the galaxy and AGN templates, 235 stellar templates from the LePhare distribution\footnote{See \url{http://www.oamp.fr/people/arnouts/LE_PHARE.html}.} were used to identify likely stars. The templates used include the set of templates from the \citet{pick98} library; the white dwarf templates of \citet{bohl95}; and the low-mass stellar templates of \citet{chab00}. In addition, a set of 5 interpolated templates was created between each of the original stellar templates, to produce a final set of 1405 stellar templates that should represent the full range of likely SEDs. 

\subsection{Photometric Redshifts}\label{S:photoz_cat}
The Zurich Extragalactic Bayesian Redshift Analyzer \citep[ZEBRA;][]{feld06} was used for the photometric redshift derivation. Default values were used for most parameters. The reader is referred to documentation included with the catalog for futher details. In the \mbox{E-CDF-S}, photometric redshifts were obtained for a total of 100,318 sources (5507 of the original 105,825 sources had detections in fewer than 3 optical bands and were not fit). Of these, 1957 are identified as stars, either photometrically or spectroscopically (including 26 white dwarfs). The remaining 97,712 sources and 649 \mbox{X-ray} AGNs were fit best by galaxy and AGN/galaxy hybrid templates, respectively. In the CDF-N, 47,224 sources and 308 \mbox{X-ray} AGNs were fit best by the galaxy and AGN/galaxy hybrid templates, and 1323 were identified as stars (including 6 white dwarfs). Tables~\ref{T:ECDFS_photoz_cat} and \ref{T:CDFN_photoz_cat} give the derived photometric redshifts, the available spectroscopic redshifts, and the photometry used by ZEBRA for these sources.
 
To assess the quality of the photometric redshifts derived by ZEBRA, comparisons are made to secure spectroscopic redshifts (given Tables \ref{T:ECDFS_photoz_cat} and \ref{T:CDFN_photoz_cat}). We used a number of quantities to assess the quality of the photometric redshifts derived by ZEBRA: the normalized median absolute deviation \citep{maro06}, $\sigma_{\rm NMAD}=1.48\times {\rm median}\left( |\Delta z - {\rm median}(\Delta z)|/[1+z_{\rm spec}] \right),$
which gives an indication of the quality of the photometric redshifts after the exclusion of outliers \citep{bram08}; the average absolute scatter, $AAS={\rm mean}\left(|\Delta z|/[1+z_{\rm spec}]\right),$
which includes the effects of outliers; and the percentage of outliers with $\frac{|\Delta z|}{1+z_{\rm spec}} > 0.1$ and $\frac{|\Delta z|}{1+z_{\rm spec}} > 0.2$, where $\Delta z=z_{\rm phot}-z_{\rm spec}$. Table \ref{T:stats} gives the quantities defined above for the photometric redshifts derived by ZEBRA for a number of subsamples.

We note that, although the above indicators are commonly used to assess the quality of photometric redshifts, the implicit assumption in their interpretation is that the spectroscopic subsample is representative of the full sample. This assumption is unlikely to be entirely true, particularly when the spectroscopic sample is small relative to the number of sources in the total sample or, as is often the case, is brighter on average than the total sample. Additionally, the template-improvement step used in our derivation introduces a bias, as we have optimized the templates for the spectroscopic subsample. The spectroscopic subsample will therefore likely have significantly better quality than the full sample (unless, again, the spectroscopic subsample is fully representative). To assess the importance of these effects, we carried out ``blind'' tests for each of the three \mbox{E-CDF-S} subsamples (the bright galaxy sample, the faint galaxy sample, and the \mbox{X-ray} AGN sample) as follows. For each subsample, we randomly used $\approx 3/4$ of the spectroscopic sources for the training procedure described above and used the remaining $\approx 1/4$ of the spectroscopic sources to test the quality of the resulting photometric redshifts. This process was repeated eight times for each subsample to ensure statistically meaningful source numbers for the test sample. The results of these blind tests (see Table~\ref{T:stats}) give the fairest assessment of the overall quality of the photometric redshifts. In general, it appears that the use of fully trained subsamples gives values for $AAS$, $\sigma_{\rm NMAD}$, and outlier fractions that are biased low by a factor of $\sim 2$--3.

\begin{deluxetable}{cccccccccccc}
\tabletypesize{\scriptsize}
\tablewidth{0pt}
\setlength{\tabcolsep}{0.03in}
\tablecaption{Photometric redshift catalog for the E-CDF-S. \label{T:ECDFS_photoz_cat}}
\tablehead{
\colhead{RA} & \colhead{Dec} & \colhead{$z_{\rm phot}$} & \colhead{$z_{\rm phot}^{\rm low~68\%}$} & \colhead{$z_{\rm phot}^{\rm up~68\%}$} & \colhead{$z_{\rm phot}^{\rm low~95\%}$} & \colhead{$z_{\rm phot}^{\rm up~95\%}$} & \colhead{$z_{\rm spec}$} & \colhead{Ref.} & \colhead{Template} & \colhead{\mbox{X-ray} ID} & \colhead{GOODS flag} \\
\colhead{(1)} & \colhead{(2)} & \colhead{(3)} & \colhead{(4)} & \colhead{(5)} & \colhead{(6)} & \colhead{(7)} & \colhead{(8)} & \colhead{(9)} & \colhead{(10)} & \colhead{(11)} & \colhead{(12)}}
\startdata
  53.0203740  & -27.7246863 & 0.105  & 0.105  & 0.123  & 0.105  & 0.172  & -1.0     &  -1    & Galaxy & -1 & 0 \\
  53.0203700  & -27.7045750 & 0.728  & 0.715 &  0.732 &  0.694  &  0.740   &  0.735 &    17    & Galaxy & -1 & 0  \\
  53.0203695  & -27.5503508 & 1.271  & 0.768 &  1.558  &  0.187 &  2.525  &  -1.0   &    -1   &  Galaxy & -1 & 0  \\
  53.0203590  &-27.7484474 & 0.956  & 0.899 &  0.996 &  0.858  & 1.064   &  -1.0    &   -1    & Galaxy & -1 & 0 
\enddata
\tablecomments{Table~\ref{T:ECDFS_photoz_cat} is presented in its entirety in the electronic edition. An abbreviated version of the table is shown here for guidance as to its form and content. The full table contains 96 columns as follows. Cols.~(1--2): Source position in degrees, Col.~(3): photometric redshift, Cols.~(4--7): estimate of the 68\% and 95\% confidence intervals of the photometric redshift, Col.~(8): spectroscopic redshift (if available), Col.~(9): source of the spectroscopic redshift (numbers correspond to those given in the references for this Table), Col.~(10): type of the best-fit template, Col.~(11): ID of the associated \mbox{X-ray} source (if any) from the 2~Ms \mbox{CDF-S} catalog of \citet{luo08} or the 250~ks \mbox{E-CDF-S} catalog of \citet{lehm05}, Col.~(12): flag indicating whether the source is inside the GOODS-S region, Cols.~(13--96): the photometry used in the fit.}
\tablerefs{(1) \citealt{vanz08}; (2) \citealt{le-f04}; (3) \citealt{szok04}; (4) \citealt{croo01}; (5) \citealt{dick04}; (6) \citealt{van-04}; (7) \citealt{bunk03}; (8) \citealt{stan04}; (9) \citealt{mign05}; (10) Silverman, Mainieri, et al., in preparation; (11) \citealt{cris00}; (12) \citealt{stro04}; (13) \citealt{ravi07}; (14) \citealt{stan04}; (15) \citealt{trei09}; (16) \citealt{pope09} (VIMOS VLT low-resolution survey); (17) \citealt{pope09} (VIMOS VLT medium-resolution survey); (18) \citealt{graz06}; (19) \citealt{zhen04}.}
\end{deluxetable}

\begin{deluxetable}{cccccccccccc}
\tabletypesize{\scriptsize}
\tablewidth{0pt}
\setlength{\tabcolsep}{0.03in}
\tablecaption{Photometric redshift catalog for the CDF-N. \label{T:CDFN_photoz_cat}}
\tablehead{
\colhead{RA} & \colhead{Dec} & \colhead{$z_{\rm phot}$} & \colhead{$z_{\rm phot}^{\rm low~68\%}$} & \colhead{$z_{\rm phot}^{\rm up~68\%}$} & \colhead{$z_{\rm phot}^{\rm low~95\%}$} & \colhead{$z_{\rm phot}^{\rm up~95\%}$} & \colhead{$z_{\rm spec}$} & \colhead{Ref.} & \colhead{Template} & \colhead{\mbox{X-ray} ID} & \colhead{GOODS flag} \\
\colhead{(1)} & \colhead{(2)} & \colhead{(3)} & \colhead{(4)} & \colhead{(5)} & \colhead{(6)} & \colhead{(7)} & \colhead{(8)} & \colhead{(9)} & \colhead{(10)} & \colhead{(11)} & \colhead{(12)}}
\startdata
 189.3129730 & 62.3347588 & 2.521 & 2.406 & 2.631 & 2.296 & 2.718 &-1.0   &       -1   &    Galaxy  &   -1   &    1 \\
 189.3709259 & 62.3344383 & 0.277 & 0.227 & 0.326 & 0.182 & 0.436 & 0.277   &       1   &    Galaxy   &  -1   &    1\\
 189.4083252 & 62.3435631 & 0.072 & 0.020 & 0.111 & 0.020 & 0.138 &-1.0   &       -1   &    Galaxy   &  -1   &    1\\
 189.3067932 & 62.3343468 & 0.258 & 0.225 & 0.304 & 0.198 & 0.539 & 0.278    &      2   &    Galaxy   &  -1   &    1
\enddata
\tablecomments{Table~\ref{T:CDFN_photoz_cat} is presented in its entirety in the electronic edition. An abbreviated version of the table is shown here for guidance as to its form and content. The full table contains 47 columns as follows. Cols.~(1--2): Source position in degrees, Col.~(3): photometric redshift, Cols.~(4--7): estimate of the 68\% and 95\% confidence intervals of the photometric redshift, Col.~(8): spectroscopic redshift (if available), Col.~(9): source of the spectroscopic redshift (numbers correspond to those given in the references for this Table), Col.~(10): type of the best-fit template, Col.~(11): ID of the associated \mbox{X-ray} source (if any) from the 2~Ms \mbox{CDF-N} catalog of \citet{alex03}, Col.~(12): flag indicating whether the source is inside the GOODS-N region, Cols.~(13--47): the photometry used in the fit.}
\tablerefs{(1) \citealt{barg08}; (2) \citealt{cowi04}; (3) \citealt{wirt04}; (4) \citealt{redd06}; (5) \citealt{barg03}; (6) \citealt{trou08}; (7) \citealt{chap05}.}
\end{deluxetable}

\begin{deluxetable}{clcllll}
\tabletypesize{\scriptsize}
\tablewidth{0pt}
\tablecaption{Photometric redshift quality estimators. \label{T:stats}}
\tablehead{
\colhead{Case} & \colhead{Sample} & \colhead{\#} & \colhead{$\sigma_{\rm NMAD}$} & \colhead{$AAS$} & \colhead{$\frac{|\Delta z|}{1+z_{\rm spec}} > 0.2$} & \colhead{$\frac{|\Delta z|}{1+z_{\rm spec}} > 0.1$} }
\startdata
\cutinhead{E-CDF-S field}
(a) &   all sources                     &  2304  & 0.0116  & 0.0262  &  2.00\% &  3.04\%  \\
(b) &   $m_R \le 24$ galaxies (trained) &  1699  & 0.0118  & 0.0215  &  1.12\% &  1.88\%  \\
(c) &   $m_R \le 24$ galaxies (blind)   &  1738  & 0.0345  & 0.0501  &  3.86\% &  8.46\%  \\
(d) &   $m_R \le 24$ galaxies (blind-C17)\tablenotemark{a}&  1530  & 0.0291 [0.0244] & 0.0404 [0.0574] &  2.87\% [7.12\%]&  6.00\% [9.61\%] \\
(e) &   $m_R > 24$ galaxies (trained)   &  605   & 0.0113  & 0.0396  &  4.46\% &  6.28\%  \\
(f) &   $m_R > 24$ galaxies (blind)     &  619   & 0.0612  & 0.1142  & 14.05\% & 25.20\%  \\
(g) &   $m_R > 24$ galaxies (blind-C17)\tablenotemark{a}  &  230   & 0.0500 [0.1057] & 0.0607 [0.1811] &  3.81\% [25.22\%]& 14.62\% [40.43\%] \\
(h) &   \mbox{X-ray} AGNs (trained)            &  283   & 0.0094  & 0.0140  &  1.06\% &  2.12\%  \\
(i) &   \mbox{X-ray} AGNs (blind)              &  315   & 0.0436  & 0.0931  & 14.92\% & 24.44\%  \\
(j) &   \mbox{X-ray} AGNs (blind-C17)\tablenotemark{a}   &  217   & 0.0495 [0.0251] & 0.0835 [0.0738] & 13.52\% [11.98\%]& 22.53\% [20.28\%] \\
\cutinhead{CDF-N field (GOODS-N region)\tablenotemark{b}}
(k) &   all sources                     &  2672  & 0.0229 [0.0440]  & 0.0480 [0.0819] &  4.27\% [8.42\%]&  8.68\% [16.84\%] \\
(l) &   $m_R \le 24$ galaxies           &  1837  & 0.0227 [0.0362] & 0.0409 [0.0497] &  2.78\% [3.54\%]&  6.26\% [9.80\%] \\
(m) &   $m_R > 24$ galaxies             &  835   & 0.0234 [0.0753] & 0.0637 [0.1526] &  7.54\% [19.16\%]& 14.01\% [32.34\%] \\
(n) &   \mbox{X-ray} AGNs                      &  164   & 0.0142 [0.0760] & 0.0380 [0.1505] &  3.66\% [20.73\%]&  7.32\% [26.22\%] \\
\cutinhead{CDF-N field (non-GOODS-N region)\tablenotemark{b}}
(o) &   all sources                     &  2687  & 0.0245 [0.0440] & 0.0538 [0.0819] &  5.02\% [8.41\%]& 10.98\% [16.86\%] \\
(p) &   $m_R \le 24$ galaxies           &  1848  & 0.0256 [0.0364] & 0.0478 [0.0497] &  3.63\% [3.52\%]&  9.58\% [9.85\%] \\
(q) &   $m_R > 24$ galaxies             &  839   & 0.0226 [0.0744] & 0.0671 [0.1528] &  8.10\% [19.19\%]& 14.06\% [32.30\%] \\
(r) &   \mbox{X-ray} AGNs                      &  212   & 0.0118 [0.0804] & 0.0457 [0.1617] &  4.72\% [21.70\%]&  7.08\% [29.72\%] \\
\enddata
\tablenotetext{a}{To obtain these numbers, we used sources with spectroscopic redshifts that have photometric redshifts in both the ZEBRA and COMBO-17 catalogs. Values in brackets are derived from the COMBO-17 catalog.}
\tablenotetext{b}{Values given for the \mbox{CDF-N} are obtained from the fully trained subsamples. Values in brackets are derived from the photometric redshift catalog of Capak et al.\ (private communication) based on the photometric catalog of \citet{capa04}.}
\end{deluxetable}

Lastly, a number of other photometric redshift catalogs exist for the \mbox{E-CDF-S} and CDF-N. In Table \ref{T:stats}, we compare the quality of our redshifts to those from two widely used photometric redshift catalogs:  the COMBO-17 catalog of the \mbox{E-CDF-S} \citep{wolf04,wolf08} and the \citet{capa04} catalog of the CDF-N. It is clear that the photometric redshifts derived by ZEBRA are comparable or superior to those of the COMBO-17 and Capak et al.\ catalogs in most respects. We also note that the catalog of Capak et al.\ has an unusual deficit of sources with $2\lesssim z \lesssim 3$; such a deficit is not present in our catalog. Lastly, we note that the photometric redshift catalog of \mbox{E-CDF-S} \mbox{X-ray} sources produced by \citet{luo10}, which uses upper limits and deblended photometry when deriving photometric redshifts, supersedes the catalog presented here for \mbox{E-CDF-S} \mbox{X-ray} sources.

\end{document}